\documentclass[a4paper,10pt]{article}
\pdfoutput=1
\usepackage{jheppub}
\usepackage[english]{babel}
\usepackage{amsmath, amssymb}
\usepackage{graphicx}
\usepackage{subfigure}
\usepackage{color}
\usepackage{xspace}
\usepackage{multirow}

\graphicspath{{./figs/}}

\newcommand{\pythia}{{\ttfamily PYTHIA}\xspace}
\newcommand{\herwig}{{\ttfamily HERWIG}\xspace}
\newcommand{\herwigpp}{{\ttfamily HERWIG++}\xspace}

\abstract{
We study the differences in the gamma--ray spectra simulated by four Monte Carlo event generator packages developed in particle physics.
Two different versions of \pythia and two of \herwig are analyzed, namely \pythia 6.418 
and \herwig 6.5.10 
in Fortran and \pythia 8.165 
and \herwig 2.6.1 
in C++. 
For all the studied channels, the intrinsic differences between them are shown to be significative and may play an important role in misunderstanding dark matter signals.
}

\title{Reliability of Monte Carlo event generators for gamma--ray dark matter searches}

\author[a]{J.~A.~R.~Cembranos}
\author[b,c]{A.~de~la~Cruz-Dombriz}
\author[a]{V.~Gammaldi}
\author[d]{R.~A.~Lineros}
\author[a]{A.~L.~Maroto}

\affiliation[a]{Departamento de F\'{\i}sica Te\'orica I, Universidad Complutense de Madrid, E-28040 Madrid, Spain}
\affiliation[b]{Instituto de Ciencias del Espacio (ICE/CSIC) and Institut d'Estudis Espacials de Catalunya (IEEC), Campus UAB, Facultat de Ci\`{e}ncies, Torre C5-Par-2a, 08193 Bellaterra (Barcelona) Spain}
\affiliation[c]{Astrophysics, Cosmology and Gravity Centre (ACGC) and Department of Mathematics and Applied Mathematics, University of Cape Town, Rondebosch 7701, Cape Town, South Africa}
\affiliation[d]{Instituto de F\'{\i}sica Corpuscular (CSIC-Universitat de Val\`{e}ncia), Apdo. 22085, E-46071 Valencia, Spain}

\begin{document}
\maketitle

\section{Introduction}
\label{1}
In the last years, numerous evidences about the existence of a new kind of invisible matter have appeared. Most of them rely on gravitational effects on galactic and extragalactic scales, such as the rotation curves of spiral galaxies, spatial distribution of gravitational lensing signals and constraints from
cosmic microwave background, among others. In spite of them, a conclusive identification of this dark component of matter has not yet been found. Although there are many plausible origins for this component~\cite{DM}, dark matter (DM) is usually assumed to be in the form of thermal relics that naturally freeze-out with the right abundance in many extensions of the Standard Model (SM) of particles~\cite{WIMPs}.
In order to confirm its nature, DM searches have followed different directions. On the one hand, DM particles can be produced in laboratory experiments such as high-energy particle colliders~\cite{Coll}.  On the other hand, local DM can be detected in a direct or indirect way~\cite{isearches,simu,Cembranos:2012nj, branonsgamma, Ce10, HessGC}.

Direct detection experiments typically operate in deep underground laboratories, while the indirect ones focus on astronomical and cosmological signal detection, with both ground based Cerenkov detectors (such as CTA, HESS and MAGIC amongst others) and satellite experiments (e.g. FERMI, PAMELA, PLANCK and WMAP). If DM particles annihilate or decay into SM particles, the signature of the final products of such processes may be detected up to some uncertainty in the astrophysical background component.
In order to set constraints on the diverse DM models
and get 
a better understanding of the
astrophysical factor associated with the distribution of this kind of matter,
numerous signals detected in gamma--rays, neutrinos, positrons, antiprotons and other particles have been studied in the available literature
\cite{Fermi, CANG, VER, HESS, Aha, MAG, WMAP,CTA,str}. Most of these analysis make use of Monte Carlo event generator packages, that allow to predict the spectra of final-state particles generated by DM annihilation and decays into SM particles. The most used Monte Carlo generator packages are \pythia and \herwig, both with available versions written either
in Fortran or C++.

In this paper we shall focus on the gamma--ray spectra generated by four softwares, showing how the choice of the Monte Carlo code may affect the DM search. Thus,
Section \ref{2} is devoted to illustrate the main differences between \pythia 6.418 (Fortran version), \pythia 8.165 (C++ version), \herwig Fortran version 6.5.10 and \herwig C++ version 2.6.1. In Section \ref{3} we determine the differences between the four Monte Carlo codes when four illustrative annihilation channels are studied.
In Section \ref{4} we then analyze the implications that these differences  may have in the WIMPs phenomenology and DM indirect searches.
Finally Section \ref{5} shall cover the main conclusions of this communication.

\section{Monte Carlo Parton Shower}
\label{2}

The differential photon flux produced by Monte Carlo event generators software may be understood as the outcome obtained from a particle shower schematization in three fundamentals parts: the QCD Final-State Radiation, the hadronization model and the QED Final-State Radiation.
Differences between available generators in the aforementioned parts, may help understanding the origin of such differences. Therefore, let us study separately the technicalities of each part as follows (read \cite{Seymour:2013ega} for further details):

\subsection{QCD Final-State Radiation}

The QCD Final-State Radiation is described by the elementary probability to radiate either quarks or gluons (partons). This probability is universal in the soft (low energy) and collinear (high energy) approximation.
In these two limits the branching probability is proportional to ~\cite{Beringer:1900zz}:
\begin{equation}
\frac{\alpha_s(k_T)}{2\pi}\Delta_s(Q_{max}^2,Q^2)P_{i,jk}(z)\frac{{\rm d}Q^2}{Q^2}{\rm d}z\frac{{\rm d}\phi}{2\pi}\;,
\label{QCD}
\end{equation}
 where $\alpha_s$ is the coupling constant of the strong interaction, $Q^2$ is the evolution variable, $Q_{max}^2$
 is its maximum allowed value, $z$ and (1- $z$) are the energy fraction of the two generated partons, and $\phi$ is the azimuthal angle ($z$ and $\phi$ are defined in the center of mass frame, but other definitions only differ beyond the leading logarithmic order approximation). $P_{i,jk}(z)$ is the Altarelli-Parisi \cite{Altarelli:1977zs} splitting function describing the distribution of the fraction $z$ of the emitted parton energy with respect to its parents, where the suffixes $i$ and $jk$ stand for the incoming and final parton species. $\Delta_s(Q^2_1,Q^2_2)$ holds for the Sudakov form factor accounting for all the non-resolvable effects of the perturbative theory (quantum loop and resonance among others) acting on the probability of transition between $Q_1$ and $Q_2$ states. $Q^2_{max}$ is set by the hard-scattering, i.e., the head (initial) process of the parton shower, and $Q_0^2$ is the last process when the parton shower ends and the hadronization begins.

The evolution variable $Q^2$ represents the first difference between the Monte Carlo simulations: In \herwig and \herwigpp $Q^2\simeq E^2(1-{\rm cos} \theta)$, where $E$ is the energy of the parent parton and $\theta$ is the emission angle. It was originally implemented in \cite{81}.
However, in \pythia 6.4 the evolution variable $Q^2$ corresponds to the virtuality of the emitted parton, i.e., its virtual mass, whereas in \pythia 8 is given by $k_T$, the transverse momentum of the emitted parton with respect to the emitting one. The latter formulation
allows to order the final-state showers with regard to $k_T$ through a sequence of falling transverse-momentum values \cite{83}.
In most cases, the two variables used in the two versions of \pythia are compatible, but \herwig turns out to reproduce more accurately the color coherence dependent data in the soft limit.

Finally, the Sudakov form factor for one parton is given by \cite{simu}:
\begin{equation}
\Delta_S(Q^2_{max},Q^2)={\rm exp}\left[-\int^{Q^2_{max}}_{Q^2}\frac{{\rm d}k^2}{k^2}\int^{z_{max}}_{z_{min}}{\rm d}z\frac{\alpha_s(z,k^2)}{2\pi}P_{i,jk}(z)\right]\;.
\end{equation}

In multiparton processes, the previous equation needs to be integrated; the integration method differs for each package. For instance, in \pythia $z_{min}=Q_0^2/Q^2$, whereas in
\herwig $z_{min}=Q_{0}/Q$. With regard to $z_{max}$, it satisfies 
$z_{max}=1-z_{min}$ for all the codes. This definition leads to conclude that, for a given value for $Q^2$,  the evolution range in the $z$ variable is larger  in \pythia  than in \herwig.
%
When comparing the two simulations with LEP data, the strong coupling constant $\alpha_s$ takes also
different values, being $\alpha_s(M_Z)\simeq0.127$ in \pythia and $\alpha_s(M_Z)\simeq0.116$ in \herwig.
This fact depends on the implemented approximation. In the QCD shower, the soft gluons interference effects lead to an ordering of subsequent emissions in terms of decreasing angles. This approximation
of coherence effects also depends on the $Q^2$ definition. For the first mass-ordering version of \pythia, in which $Q^2\approx m^2$ with $m^2=E^2-k^2\geq 0$, it had to be implemented as additional requirement. In the case of the $k_T$-ordering version, with $Q^2\approx k_T^2=z(1-z)m^2$, it leads directly to the proper behavior. Finally, due to theoretical analysis, the scale choice $\alpha_s=\alpha_s(k_T^2)=\alpha_s(z(1-z)m^2)$ is the default one in \pythia. On the other hand, \herwig takes into account this effect via the angular ordering of emissions in the parton shower by redefining the running constant. 
In this case, $\alpha_s=\alpha_s\left(z^2(1-z^2)\tilde q^2\right)$, where $\tilde q$ corresponds to the scale of the decaying parton. Moreover, a two-loop approximation is reproduced in \herwig by means of the Monte Carlo scheme with $\alpha_s^{MC}=\alpha_s^{\bar{MS}}(1+K\alpha_s^{\bar{MS}}/2\pi)$, where $\alpha_s^{\bar{MS}}$ is defined in the usual modified minimal subtraction ($\bar{MS}$) scheme in QCD (read \cite{CMW} for further details). In any case, we conclude that photon emission is not affected by angular ordering \cite{Py6}.
\subsection{Hadronization}

When the evolution variable $Q^2$ reaches the value $Q_0^2$, the parton shower ends and the hadronization begins. Two different models to describe hadronization are thus developed in the two aforementioned packages. \pythia relies on the String Model Hadronization~\cite{Py6, Py8} whereas \herwig does on the Cluster Model Hadronization~\cite{Her, Her++}.
In any case, both models take into account the experimental data collected by the LEP for tuning their parameters. In particular, the standard ``tunes'' use data at 100~GeV of center of energy.
In the future, new tunes could also consider 
the LHC data.
In any case, the hadronization model does not seem to affect the gamma--ray spectra in an appreciable way, except if the $\pi^0$ production changes significantly.
Finally, let us remember that most of the hadrons formed during the hadronization process are unstable and will eventually decay.
The resultant final states, which are mainly leptons, lead the photon production involving QED processes.

\subsection{QED Final-State Radiation}

The radiation emitted by quarks, $W^{\pm}$ bosons, and charged leptons (i.e. Bremsstrahlung radiation), as well as the possibility of pair production, can be added to equation (\ref{QCD}) introduced above.
The Bremsstrahlung component of the Final-State Radiation (FSR) represents the main contribution in the case of gamma--rays produced by DM annihilating/decaying into $e^+e^-$ and $\mu^+\mu^-$ channels.
The high energy leptons come directly from the hard process in the first case and both from hard processes and $\mu^{\pm}$ decay in the second one. In any case, associated $\gamma$-photons are produced by Bremsstrahlung effects in both cases.
Bremsstrahlung FSR from hard processes is currently not implemented in \herwigpp version 2.6.1, being unable to produce gamma--ray spectra in the case of $e^+e^-$ and $\mu^+\mu^-$ channels,
while it is included in both \herwig and \pythia (6.4 and 8).
This component clearly affects all the logarithmic part of gamma--ray spectra at high energy generated with \herwigpp, as shall be shown in the following sections.\\
%

With regards to the electroweak (EW) $2\rightarrow2$ processes of the FSR, where photons are produced or annihilated, \pythia 8 accounts for all these processes, except the $\gamma\gamma \rightarrow W^+W^-$. As for \herwig, it contains the $q \to q\gamma$ processes, but not the process $\gamma\to f\bar f$.
These two last processes are indeed contained in \herwigpp. However, we verified that different sets of such processes did not affect the gamma--ray spectra in an appreciable way after modifying the codes.


\section{Gamma--ray spectra from dark matter annihilation/decay}
\label{3}
In this Section we study the spectra of four relevant channels by using the four Monte Carlo generators mentioned above.
Namely, we have studied the on-shell channels: $W^+W^-$, $b\bar{b}$, $\tau^{+}\tau^{-}$ and $t\bar{t}$ since they are representative channels of the phenomenology of annihiliating/decaying DM. 
The $t \bar{t}$ channel was studied separately since it presents a particular phenomenology with respect to the other quark channels.

The photon spectra is better described in terms of the dimensionless variable:
\begin{equation}
	x \equiv 2 \frac{E_{\gamma}}{E_{\rm CM}} \, ,
\end{equation}
where $E_{\gamma}$ and $E_{\rm CM}$ correspond to the photon and center of mass (CM) energies, respectively. This variable is simply reduced to $x \equiv E_{\gamma}/M_{\rm DM}$ in the case of annihilating DM and therefore lies in the range between 0 to 1.
Large differences between spectra are usually present at extremes of $x$. For this reason, we present the spectra in both linear and logarithmic scales for $x$. In the first (second) case the
behavior at high (low) $x$ is more clearly shown.
%
For each channel, we focused on two values of DM particle mass: $100$~GeV and $1$~TeV. In the case of the $t\bar{t}$  channel the masses under study were $500$~GeV and $1$~TeV.

\subsection{Gamma--ray spectra from DM annihilation: $W^+W^-$ channel}

\begin{figure}[tb]
\centering
\resizebox{\columnwidth}{!}{
\includegraphics[height=100pt]{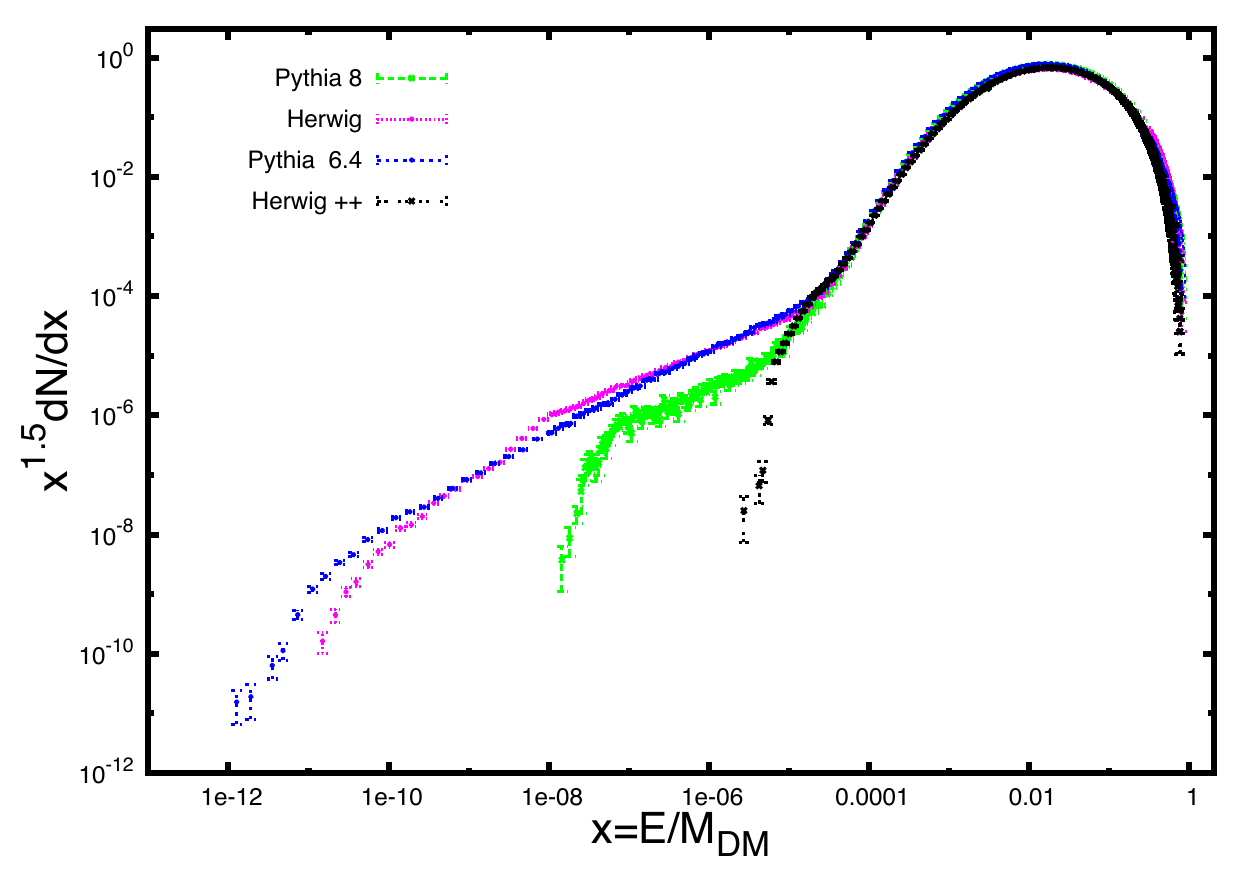}\includegraphics[height=100pt]{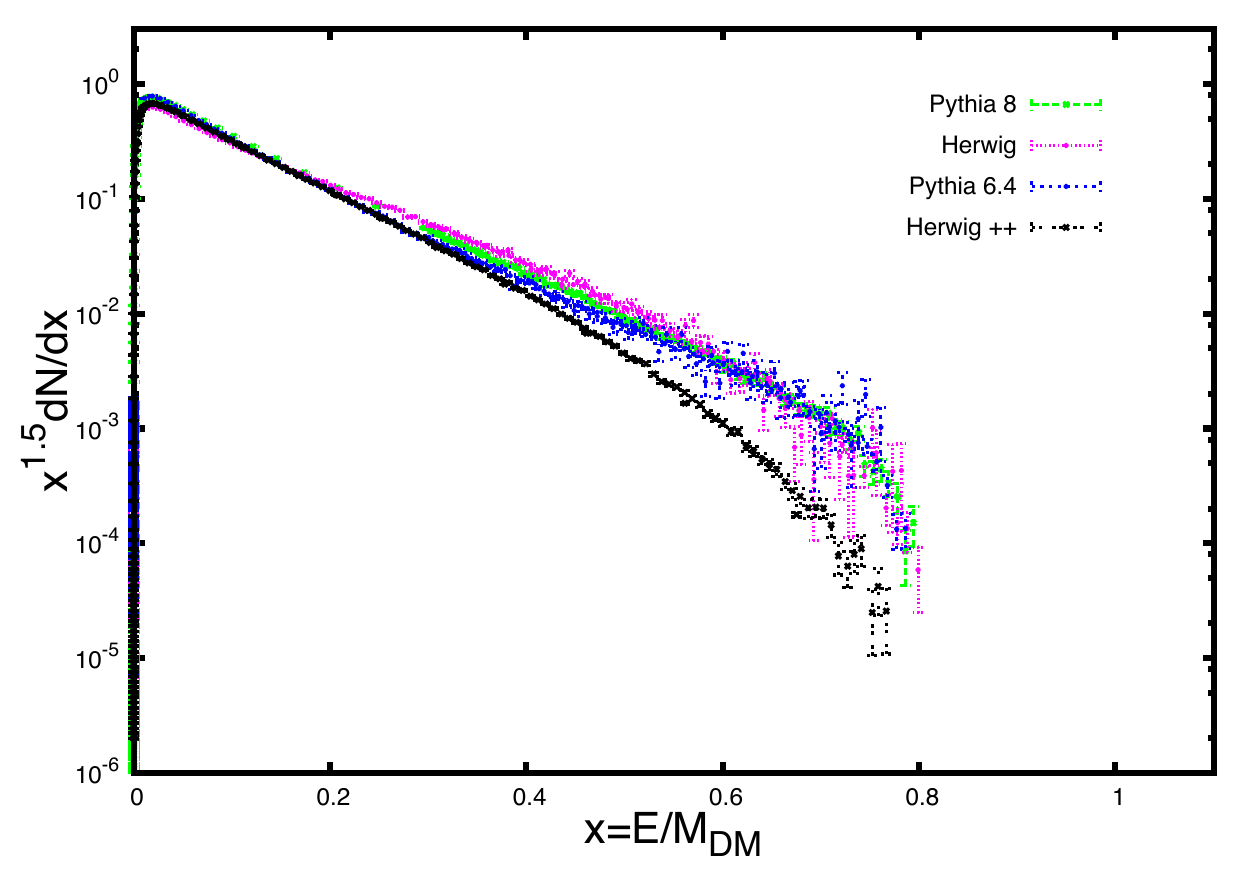}}
\caption{
{\it (Left--panel)} DM particles annihilating into $W^+W^-$  channel with $M_{\rm DM}=100$ GeV in logarithmic scale. The simulations are consistent down to $x\simeq10^{-4}$. At $x\simeq10^{-5}$ Fortran simulations are bigger than the C++ ones by a factor ten. At $x\simeq10^{-6}$ no more photons are produced in \herwigpp provided that the \texttt{QEDRadiationHandler} is set off as default. In our simulation, \texttt{QEDRadiationHandler} is switched on with a clear cut-off at energy of $10^{-10}$. Analogous cut-off appear
at $x\simeq10^{-8}$ in \pythia 8, $x\simeq10^{-11}$ in \herwig and $x\simeq10^{-12}$ in \pythia 6.4 . The simulations are very different at these energy values and physical validity has to be checked. Due to the fit of the Monte Carlo software with high energy colliders (such as LEP and LHC) that are poor of data at low energy, simulations at low energies might be unreliable. If this is the case, it is expected that this effect affects all the simulated channels.
{\it (Right--panel)} DM particles annihilating into $W^+W^-$ channel with $M_{\rm DM}=100$ GeV in linear scale.
Notice the lower flux for \herwigpp at high energies when compared to the  rest of packages
}
\label{fig:w100g}
\end{figure}

\begin{figure}[tb]
\centering
\resizebox{\columnwidth}{!}
{\includegraphics[height=100pt]{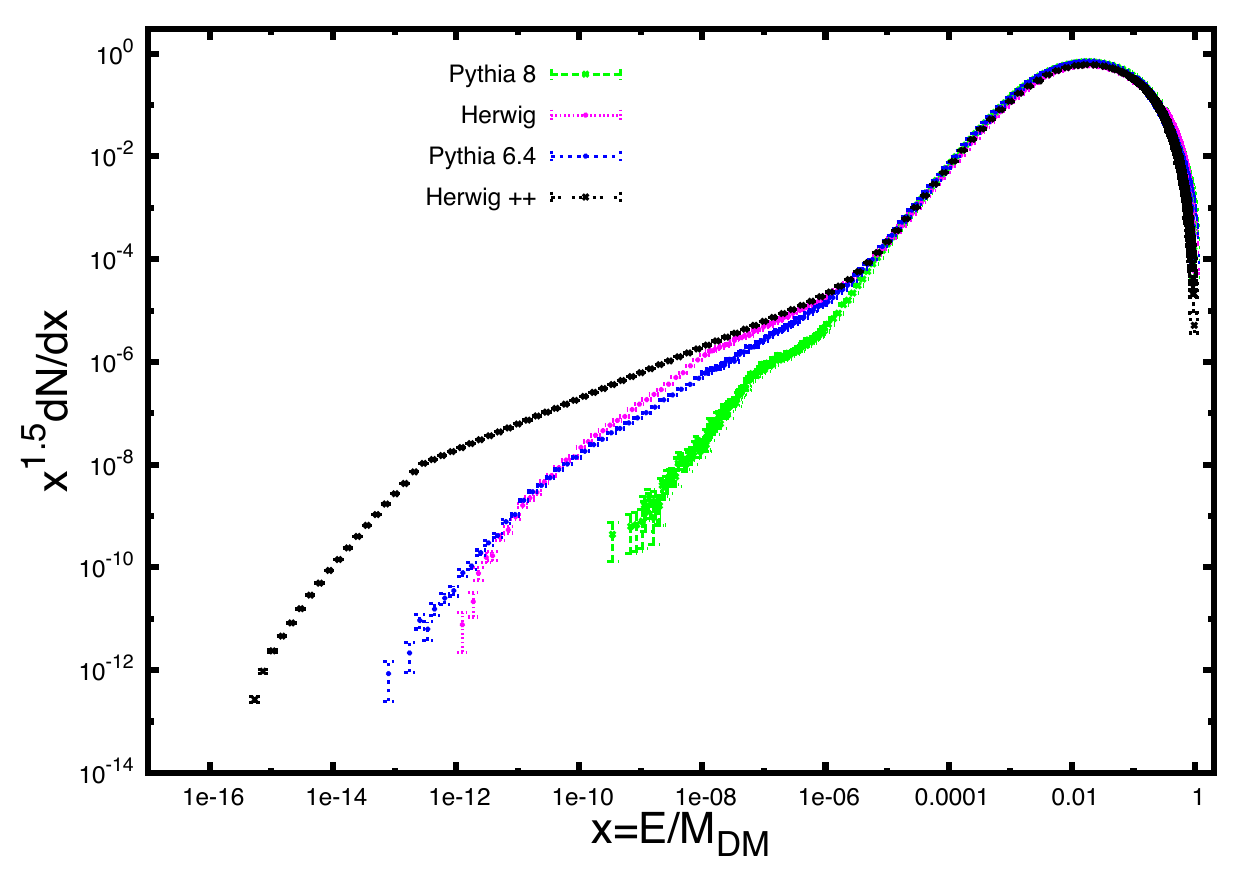}\includegraphics[height=100pt]{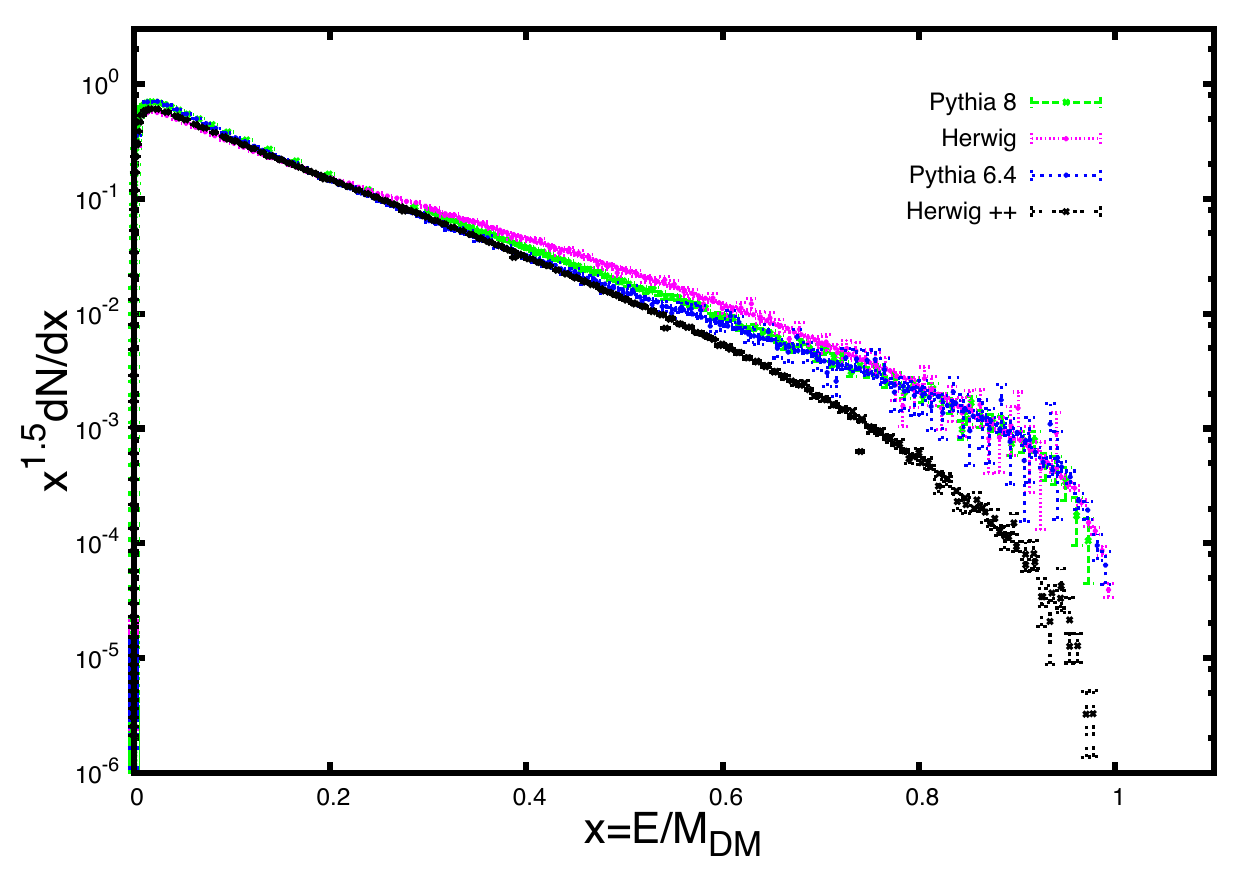}}
\caption{
{\it (Left--panel)} $W^+W^-$ annihilation channel with $M_{\rm DM}=1$ TeV in logarithmic scale. As in Fig.~\ref{fig:w100g}, the simulations are consistent down to a value of $x$, that is $10^{-6}$ in the case of $M_{\rm DM}=1$ TeV (a factor ten lower in $x$ with respect to the case with $M_{\rm DM}=100$ GeV). Similar behaviors of the lower energy cuts-off are also observed, with a general shift of $x$ cut-off value of order $10^{-2}$.
{\it (Right--panel)} $W^+W^-$ annihilation channel with $M_{\rm DM}=1$ TeV in linear scale. All the  simulations except for \herwigpp exhibit the same behavior as in Fig. \ref{fig:w100g}, but within $x\simeq0.3$ and $x\simeq0.7$ and a maximum discrepancy at $x\simeq0.5$.  The shift with respect to Fig. \ref{fig:w100g} can be simply explained by the increment of the WIMP mass.}
\label{fig:w1000g}
\end{figure}

\begin{figure}[tb]
\centering
\resizebox{\columnwidth}{!}{
\includegraphics[height=100pt]{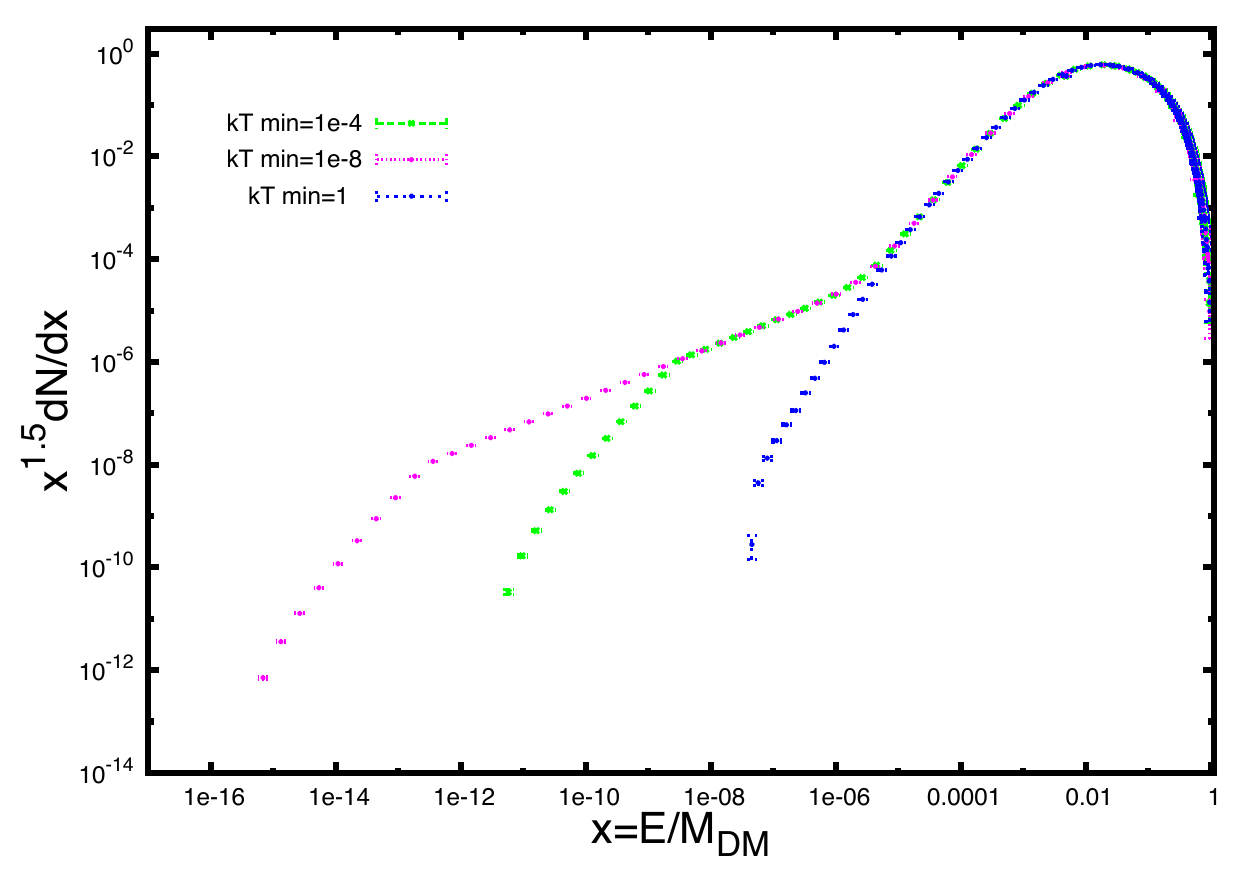}\includegraphics[height=100pt]{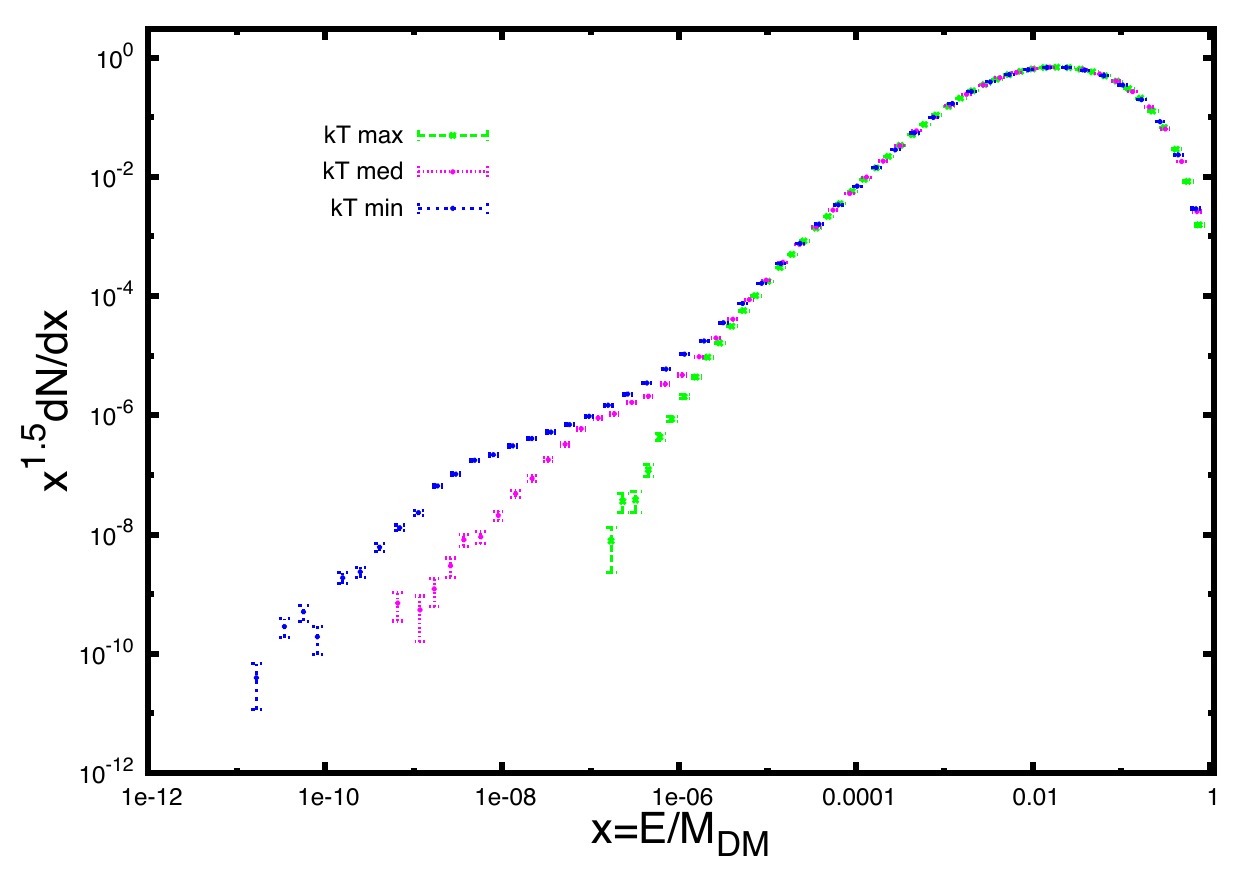}}
\caption{
Cut-off at low energy photons in C++ codes. High energy linear scale are not affected.
 {\it (Left--panel)} $W^+W^-$ annihilation channel with HERWIG++ at $M_{\rm DM}=1$ TeV in logarithmic scale. Different cut-off at low energy in logarithmic scale correspond to cuts in the \texttt{QEDRadiationHandler} of $k_T=10^{-8}\,, 10^{-4}\,, 1$. {\it (Right--panel)} $b\bar b$ annihilation channel with PYTHIA 8 at $M_{DM}=1$ TeV in logarithmic scale. Here the cut-off are set as the minimum, medium and maximum value of the allowed range of value.}
\label{fig:w1tevqed}
\end{figure}

The simulated gamma--ray spectra for DM particles annihilating into $W^+W^-$ channel appear very similar for $ x > 10^{-5}$ both for a DM mass of $100$ GeV and $1$ TeV.
This behavior can be seen in Figs. \ref{fig:w100g} and \ref{fig:w1000g} respectively.
It is clear from the figure the considerably lower fluxes generated by \herwigpp at high
energies as compared
to the rest of packages, probably because of the absence of Bremsstrahlung from hard processes
in the  $e^+e^-$ and $\mu^+\mu^-$ cases commented before.
On the other hand, a slight difference is observed for energies between $x = 0.3 - 0.7$ with \herwig providing in both cases the highest values.
Nonetheless, the main differences appear at lower energies as can be seen in Figs.~\ref{fig:w100g} and \ref{fig:w1000g}.
%
%
In \pythia 8, we have generated each photon spectrum by using the resonant process $e^+ e^- \rightarrow \phi^{*}$, where $\phi^{*}$ is a resonance with mass of $E_{\rm CM}$ and a user-defined decay mode.
This procedure is very similar to the one we used for \pythia 6.4,  except that channels were created by using the subroutine \texttt{PY2ENT}.
In \herwigpp, we used the scattering of photons as the initial process.
%
The photon spectra are then independent of the initial beams ($e^+ e^-$ or $\gamma \gamma$) and solely depend on the energy of the event, i.e. $E_{\rm CM} = 2 M_{\rm DM}$.
In \pythia 8, the cut-off at low energy strongly depends upon this parameter \texttt{pTminChgL} (dubbed here $p_T$) and exactly corresponds to its set value, with allowed range of 0.001 -- 2.0 and a default value of 0.005.
(FIG. \ref{fig:w1tevqed}  {\it (Right-panel)})
In \herwigpp, \texttt{QEDRadiationHandler} is set off by default, so that the cut-off appears to higher energy with respect to the other Monte Carlo generators.
In the opposite case, when \texttt{QEDRadiationHandler} is enable and the relevant parameter \texttt{IFDipole:MinimumEnergyRest} varies in values, the spectrum at low energy changes drastically. Smaller values of such parameter enlarge the production of photons at low energies (See Fig. \ref{fig:w1tevqed}, {\it left--panel}).

%

\subsection{Gamma--ray spectra from DM annihilation: $b\bar b$ channel}
\begin{figure}[tb]
\centering
\resizebox{\columnwidth}{!}
{\includegraphics[height=100pt]{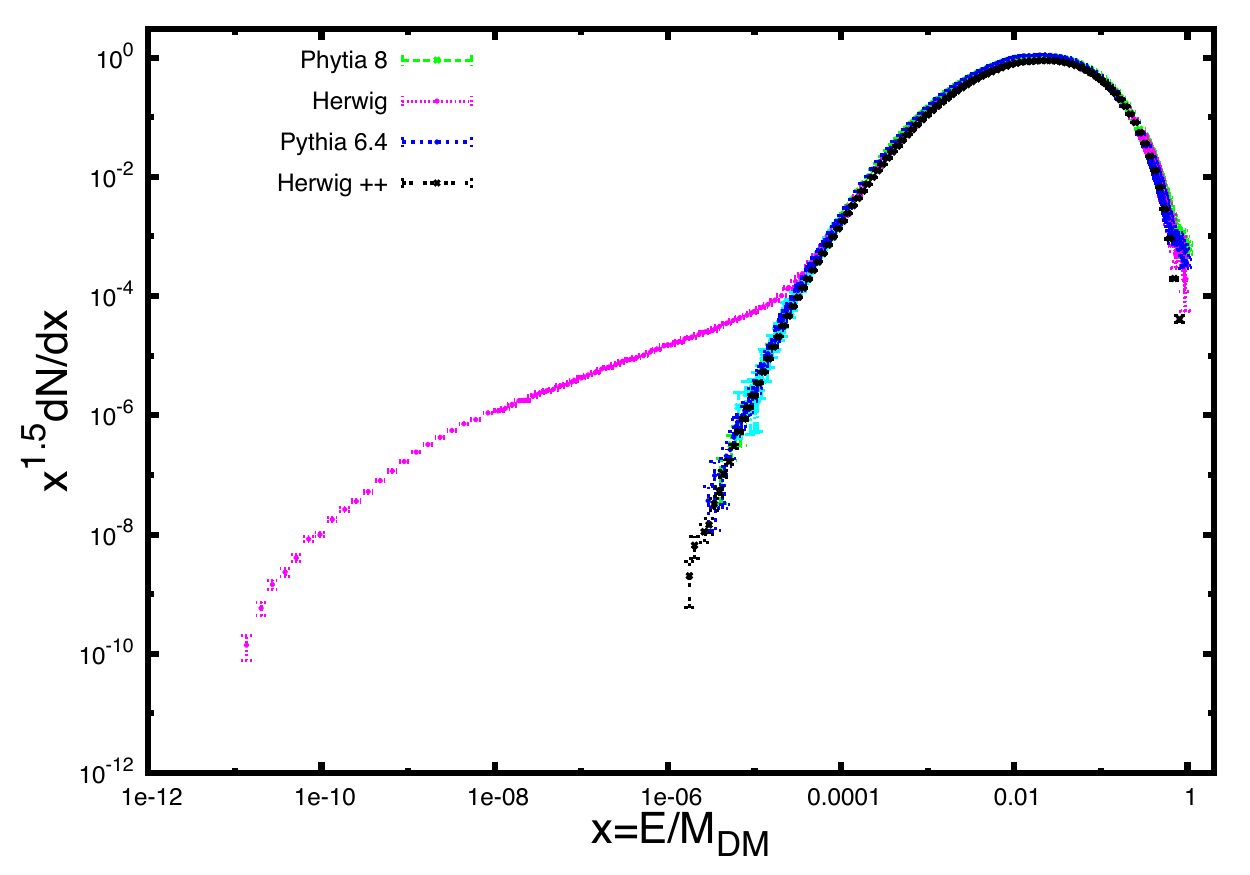}\includegraphics[height=100pt]{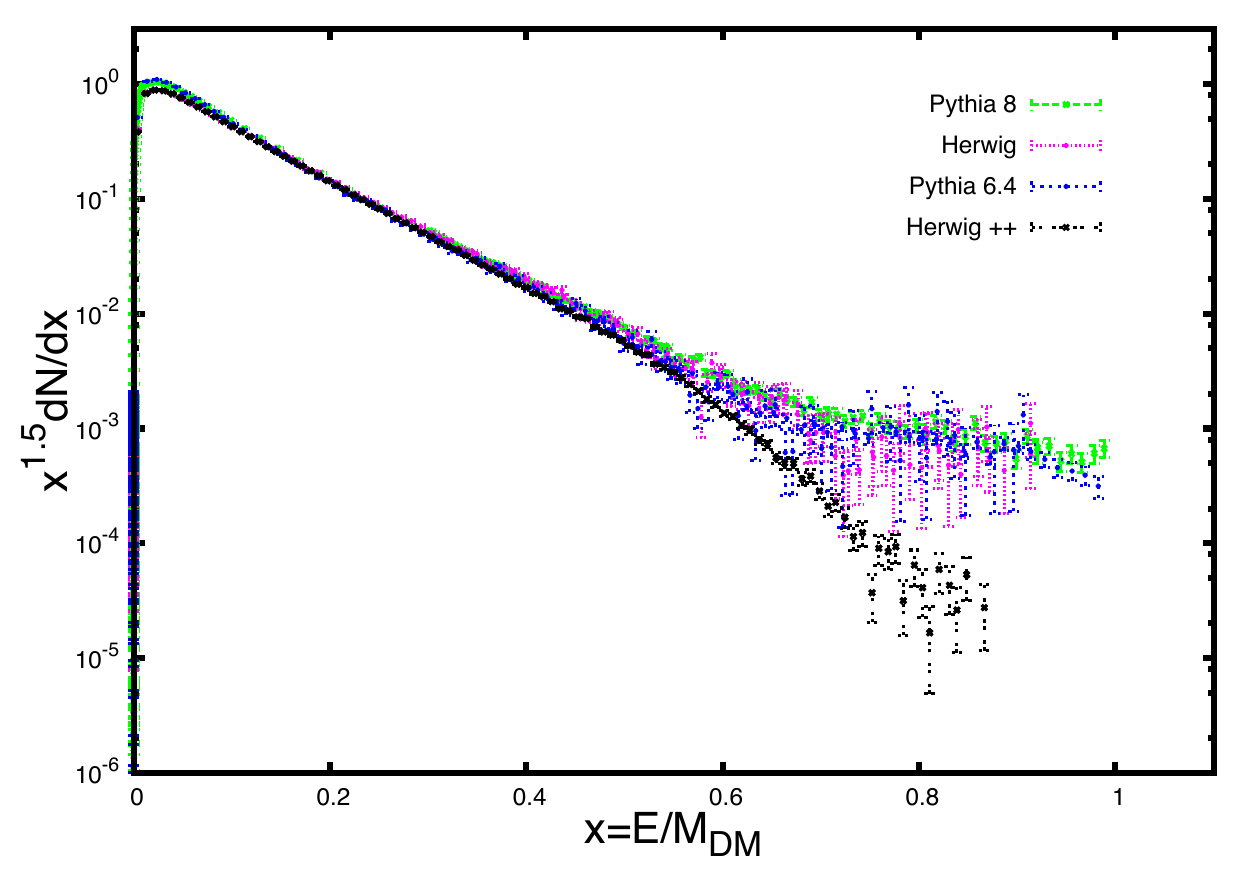}}
\caption{
{\it (Left--panel)} $b\bar b$ annihilation channel with $M_{\rm DM}=100$ GeV in logarithmic scale. Three of the four simulations perfectly match down to $x\simeq10^{-6}$, where no more photons are produced. \herwig Fortran also match down to $\simeq10^{-5}$. Here, its simulated flux appears much bigger, with no photons counted at energies smaller than $x\simeq10^{-11}$.
{\it (Right--panel)} $b \bar b$ annihilation channel with $M_{\rm DM}=100$ GeV in linear scale. Three of the four simulations are in agreement within the statistical error bars on the full $x$ range, while \herwigpp gives lower flux above  $x\simeq0.5$.
}
\label{fig:b100g}
\end{figure}

\begin{figure}[tb]
\centering
\resizebox{\columnwidth}{!}
{\includegraphics[height=100pt]{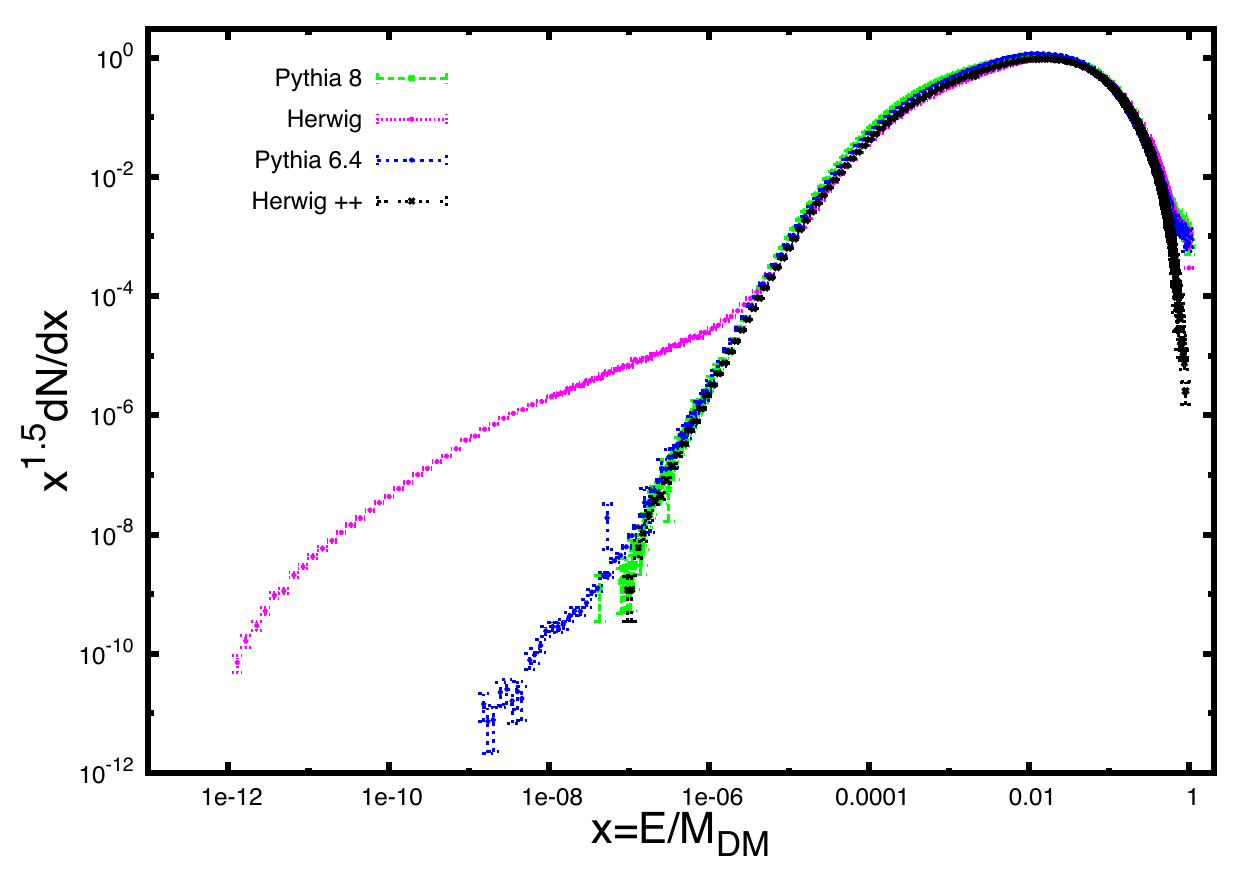} \includegraphics[height=100pt]{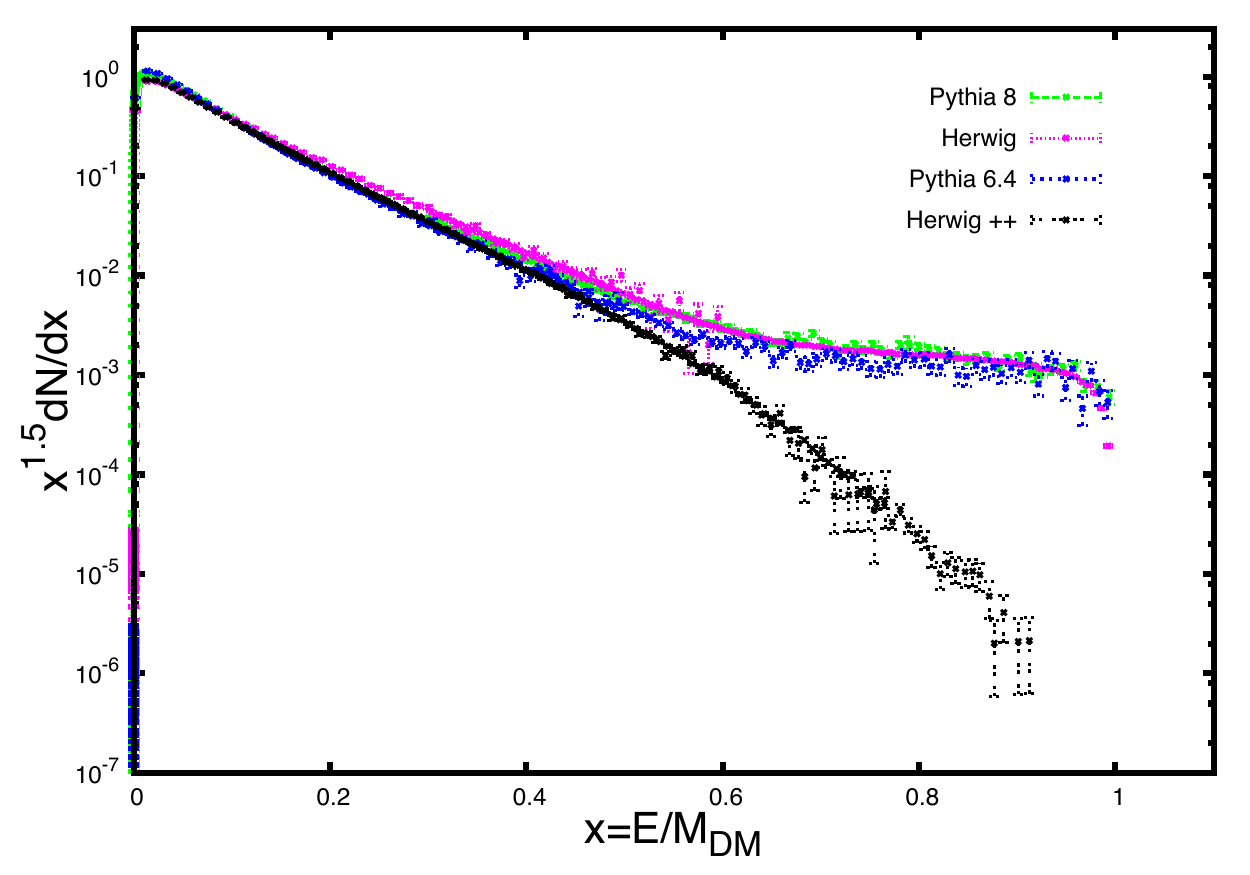}}
\caption{
{\it (Left--panel)}
$b \bar b$ annihilation channel with $M_{\rm DM}=1$ TeV in logarithmic scale. \pythia 6.4 agrees with both \herwigpp and \pythia 8 down to $x\simeq10^{-7}$, where the spectra of the latter two packages stop. \pythia 6.4 stops providing gamma--rays at $x\simeq 10^{-9}$. 
Once again, \herwig generates larger gamma--ray fluxes at low energy. The difference at high energy discussed in Fig. \ref{fig:w1000g} is also apparent on the right panel.
{\it (Right--panel)}
$b \bar b$ annihilation channel with $M_{\rm DM}=1$ TeV in linear scale. As in Fig. \ref{fig:b100g}, \herwigpp gives much lower flux above $x\simeq0.5$.
Although \herwig agrees both with Pithia 6.4 and \pythia 8 within statistical errors, \pythia 8 flux (with better statistics) appears two or three times bigger than \pythia 6.4 at $x\simeq0.6, 0.8$.}
\label{fig:b1t}
\end{figure}

In the case of DM annihilation into $b \bar{b}$ channel, the \herwigpp spectrum appears lower for high energy $\left( x > 0.6\right)$ with respect to the other simulations, due to the lack of the Bremsstrahlung photons generated by high energy leptons.
Thus, both \pythia codes and \herwig  simulations look very similar qualitatively for the two studied values of DM mass
as seen in Figs. \ref{fig:b100g} and \ref{fig:b1t}.
On the other hand, at very small energies $\left(x<10^{-4}\right)$ \herwig simulation returns higher values of the flux
with respect to the other packages. This fact can be seen in Figs. \ref{fig:b100g} and \ref{fig:b1t}. The other codes for these small energies
agree very well in their predictions.

\subsection{Gamma--ray spectra from DM annihilation: $\tau^+\tau^-$ channel}

\begin{figure}[tb]
\centering
\resizebox{\columnwidth}{!}
{\includegraphics[height=100pt]{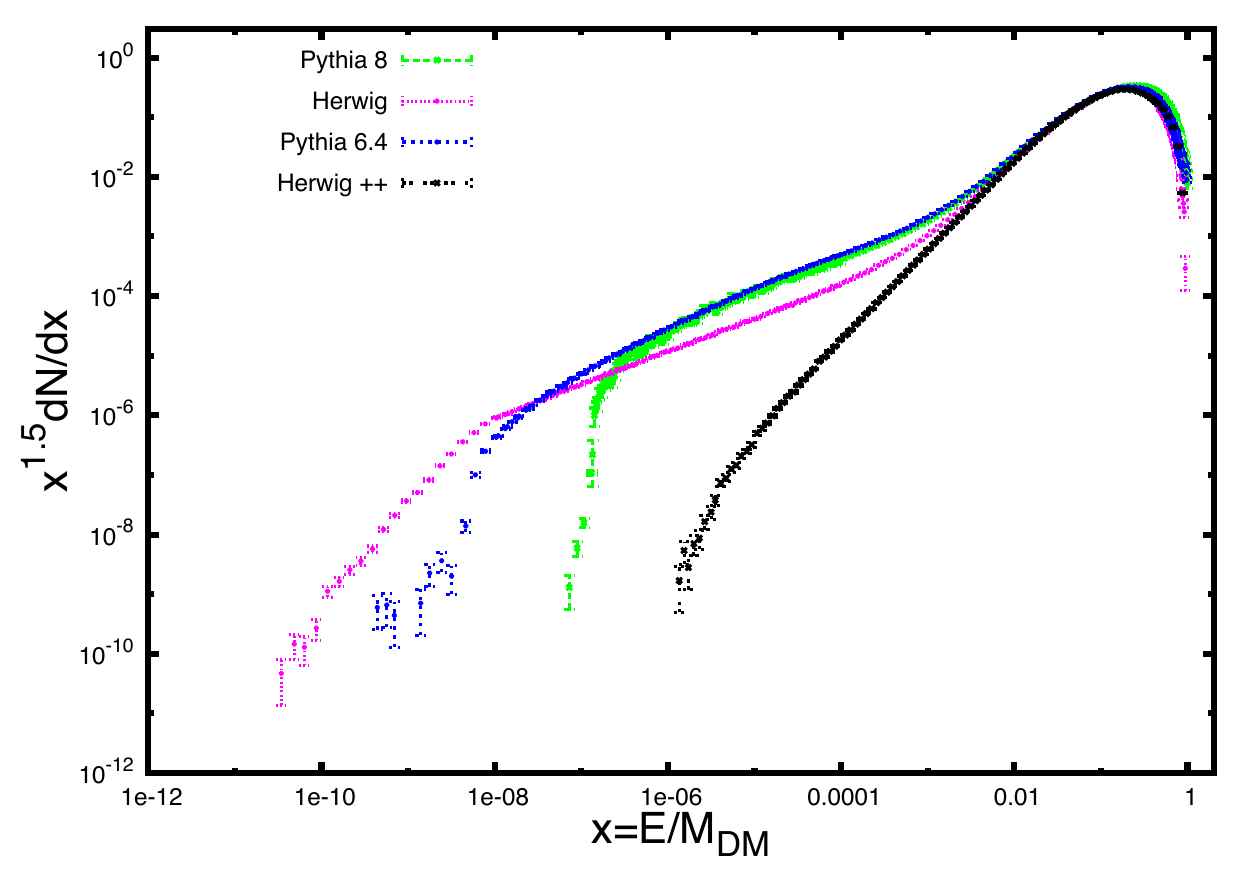}\includegraphics[height=100pt]{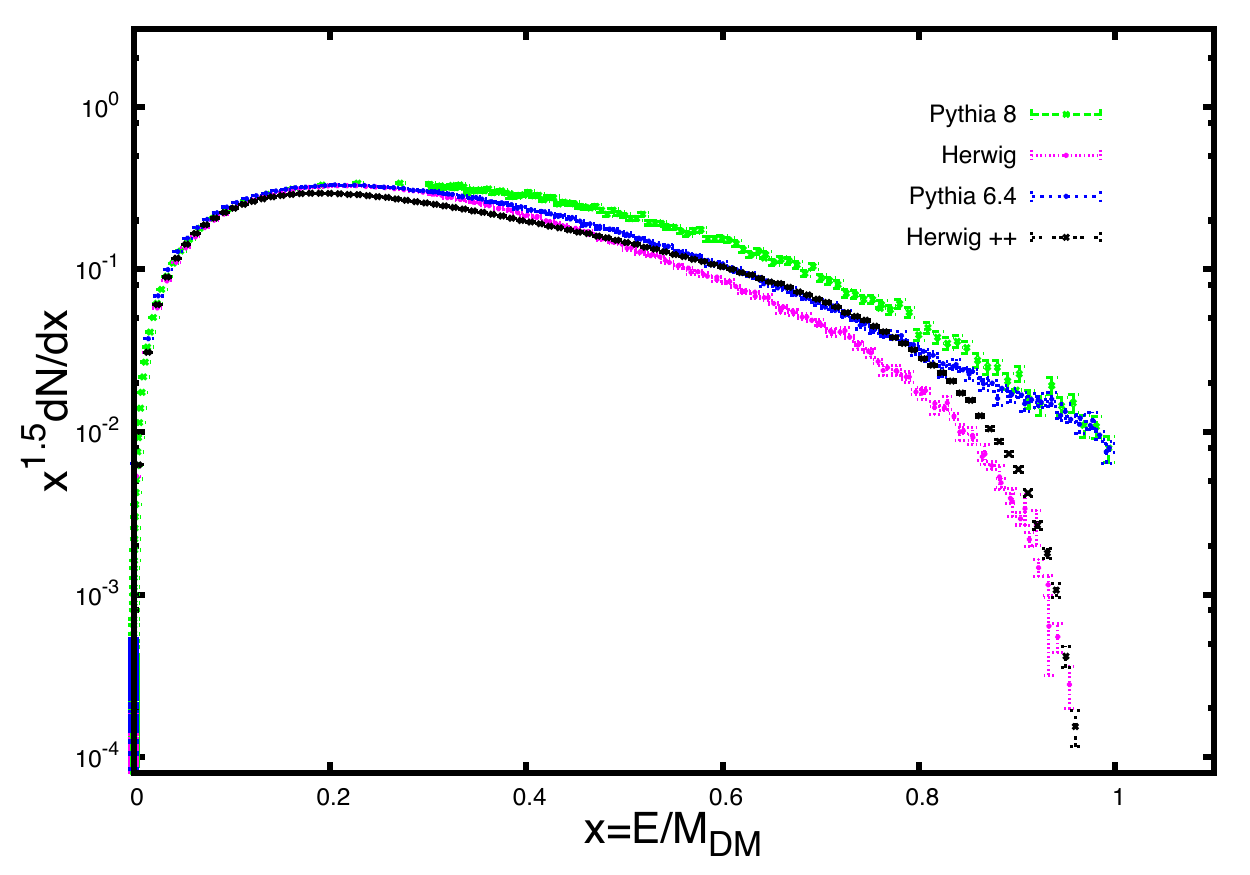}}
\caption{
{\it (Left--panel)} $\tau^+\tau^-$ annihilation channel with $M_{\rm DM}=100$ GeV in logarithmic scale. The simulations are inconsistent below $x\simeq10^{-2}$.  \pythia codes are more consistent, generating the same spectral form down to $x\simeq10^{-7}$, where \pythia 8 has its cut-off. \pythia 6.4 spectra attains smaller energies to almost $10^{-10}$.
\herwig cut-off reaches almost $x\simeq10^{-11}$, but its flux is lower than the \pythia ones below $x=10^{-3}$ and reaching the maximum inconsistence
of almost a factor ten at $x=10^{-5}$.
\herwigpp appears totally inconsistent with the other three packages, with a much lower flux that gets a maximum divergence
of 5 orders of magnitude at $x\simeq10^{-6}$ where its photons production stops.
{\it (Right--panel)} $\tau^+\tau^-$ annihilation channel with $M_{\rm DM}=100$ GeV in linear scale. For this leptonic channel, the spectral forms of the four codes differ on the whole energy range. We can see that the spectral cut-off at high energy is similar for both \herwig codes and \pythia ones by pairs. In the interval $x \simeq 0.6-0.8$, simulated gamma--ray flux
from \pythia 6.4 and \herwigpp match.
At $x\simeq0.7$, \pythia 8 lies a factor 2-3  above \herwigpp and \pythia whereas \herwig lies the same factor below. Therefore there exists a non negligible difference (almost a factor ten), between  \pythia 8 and \herwig simulated spectra at this value of $x$.}
\label{fig:tau100g}
\end{figure}

\begin{figure}[tb]
\centering
\resizebox{\columnwidth}{!}
{\includegraphics[height=100pt]{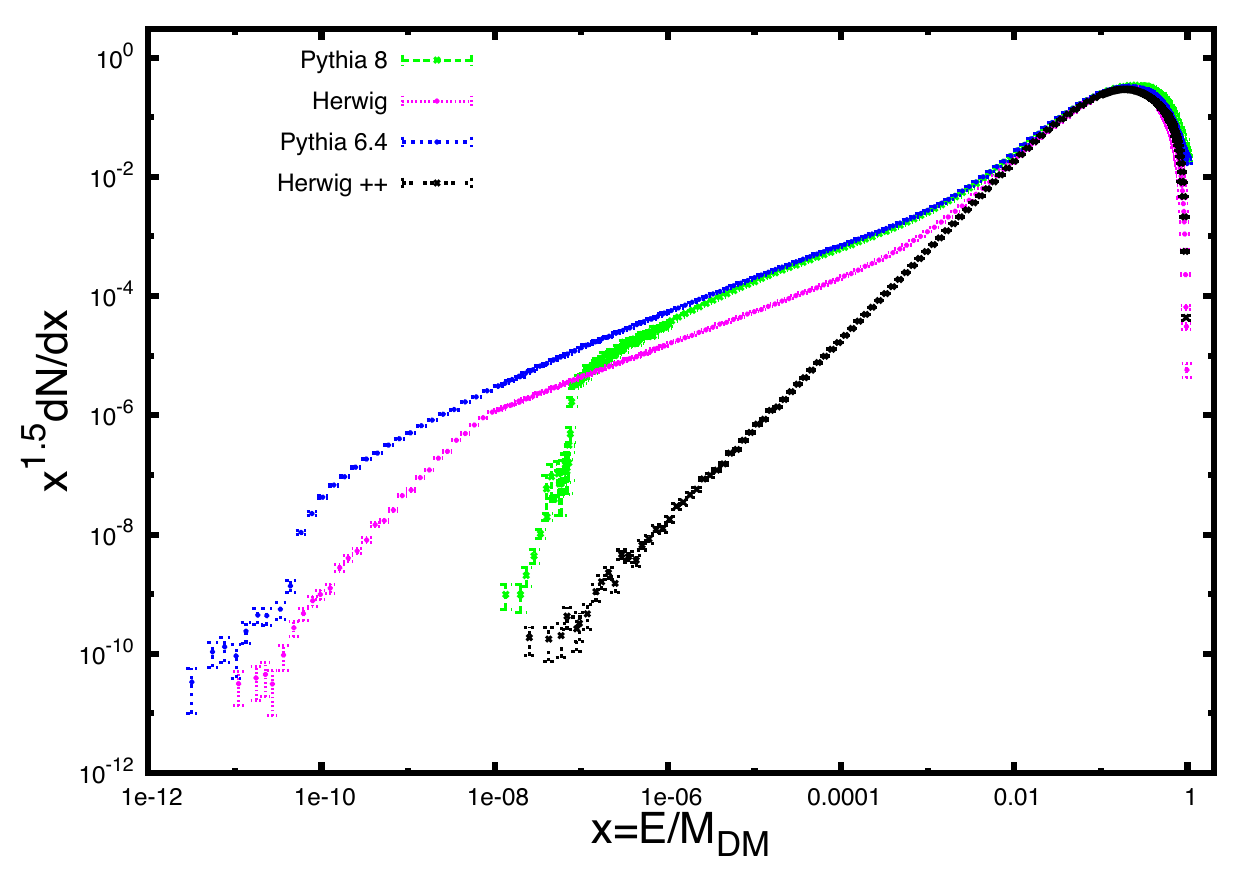}\includegraphics[height=100pt]{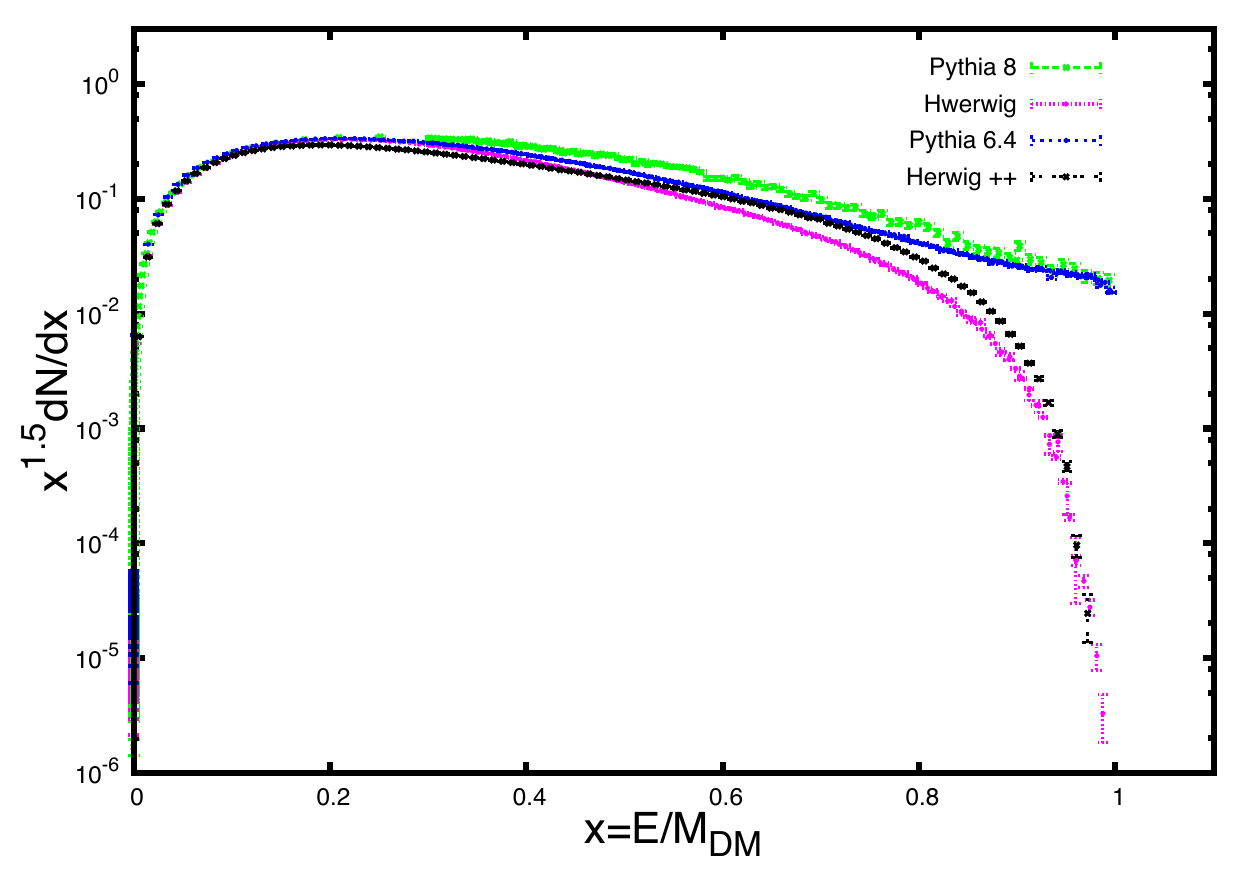}}
\caption {
{\it (Left--panel)} $\tau^+\tau^-$ annihilation channel with $M_{\rm DM}=1$ TeV in logarithmic scale. Compared to Fig. \ref{fig:tau100g}, all the lower cut-offs are shifted by a factor of ten to lower $x$'s, with the exception of \pythia 6.4 that is shifted by a factor of a hundred, so that it never crossed \herwig data as happened with $M_{\rm DM}=100$GeV.
{\it (Right--panel)} $\tau^+\tau^-$ annihilation channel with $M_{\rm DM}=1$ TeV in linear scale. The behavior is analogous to the one discussed in Fig. \ref{fig:tau100g}.}
\label{fig:tau1t}
\end{figure}

Differences in the gamma-spectra appear in the case of DM particles annihilating into leptonic channels. Here we show the $\tau^+\tau^-$ annihilation channel as an illustrative example. In this channel and for the two studied DM masses, both \herwig codes present an important suppression of the spectrum for energies in the interval $0.8<x<1$, while both versions of \pythia extend the photon spectra up to $x=1$ with higher spectra. This fact can be observed
in Figs. \ref{fig:tau100g} and \ref{fig:tau1t} and
%
may be explained by the absence of Bremsstrahlung gamma--rays generated by high energy leptons when \herwig codes are used.
As can be seen in the leptonic and muonic channel, \herwig Fortran accounts for an extrapolation with respect to the Bremsstrahlung photons related with hard processes, but it does not provide an exact implementation of this EW process. This is the reason why the gamma--ray spectra simulated with \herwig Fortran for channels where the Bremsstrahlung radiation contribution is subdominant are in agreement with \pythia 6.4 and 8 results, up to the statistical errors.
Moreover, a difference of one order of magnitude appears for energies $x \ge 0.8$ among \pythia codes and \herwig codes.
At intermediate energies, $x \approx \, 10^{-3} - 0.2$, all codes agree.
For small energies, \pythia packages agree in their spectra up to $x=10^{-7}$ but not for lower energies where both \pythia 8 seems to be strongly suppressed for energies smaller than $x=10^{-7}$. 

\herwigpp produces less photons for small energies $x \le 10^{-3}$, although the \texttt{QEDRadiationHandler} was enable.
Concerning \herwig, the spectrum can be extended down to $ x = 10^{-11}$ and it lies in between the \pythia 6.4 and \herwigpp simulations, for the two studied masses and for small energies.
With regard to high energies close to $x=1$, \herwig spectrum is the most suppressed for this channel.



\begin{figure}[tb]
\centering
\resizebox{\columnwidth}{!}
{\includegraphics[height=100pt]{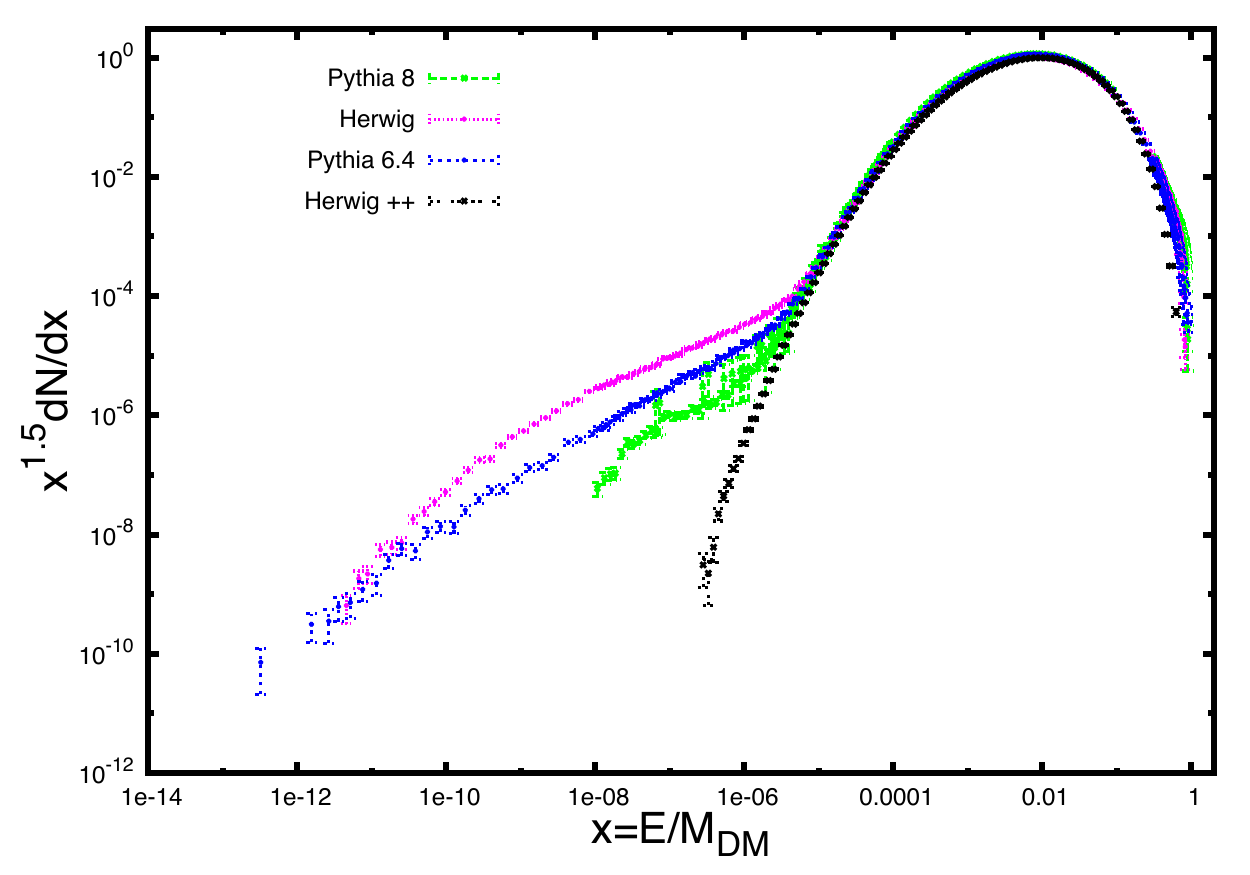}\includegraphics[height=100pt]{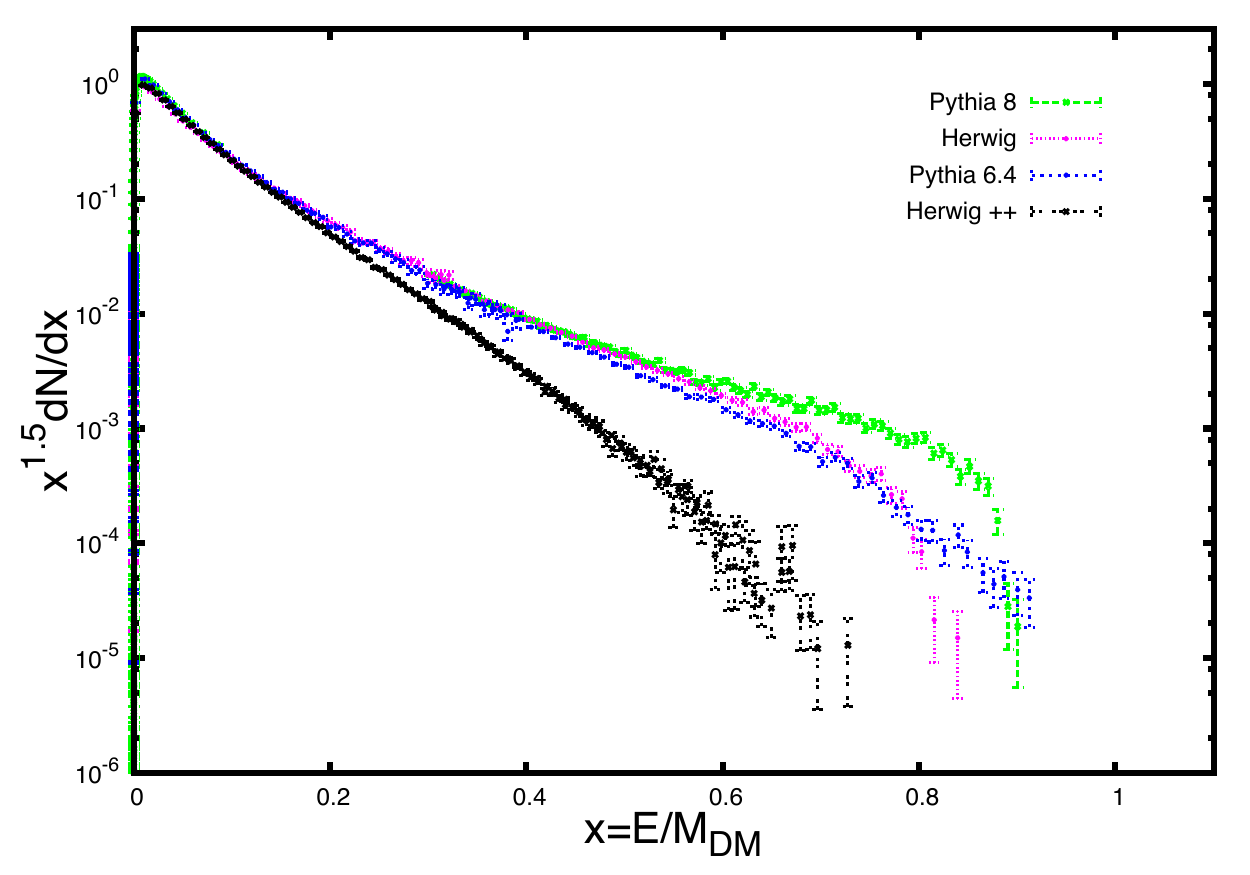}}
\caption {
{\it (Left--panel)}
$t \bar t$ annihilation channel with $M_{\rm DM}=500$ GeV in logarithmic scale. At low energy the simulations are consistent down to $x\simeq10^{-5}$. \herwigpp drops down at $x\simeq10^{-7}$ and \pythia 8 does at $10^{-9}$, producing a higher number of photons 100 times bigger than \herwigpp at $x\approx10^{-7}$, and almost 10 times lower of \pythia 6.4 at the same value of $x$. \pythia 6.4 cuts-off at $x\simeq 10^{-13}$ and 
\herwig does at  $x\simeq 10^{-12}$, where the two spectra match. For higher energies, \herwig gamma--ray flux is higher than \pythia 6.4, with  a maximum factor of ten at
$x\simeq 10^{-9}$.
{\it (Right--panel)} $t \bar t$ annihilation channel with $M_{\rm DM}=500$ GeV linear scale. The four simulations are manifestly inconsistent between them at high energy. \herwigpp flux became lower from $x\simeq0.2$ onwards and cuts off at $x<0.8$. At $x\simeq0.4$ \pythia 6.4 and \herwig are similar between the statistical errors up to $x\approx0.8$, where spectra and cuts-off become different. \pythia 8 starting from $x\simeq0.6$ produces the highest flux with cut-off at $x\simeq 1$.}
\label{fig:top5g}
\end{figure}

\begin{figure}[tb]
\centering
\resizebox{\columnwidth}{!}
{\includegraphics[height=100pt]{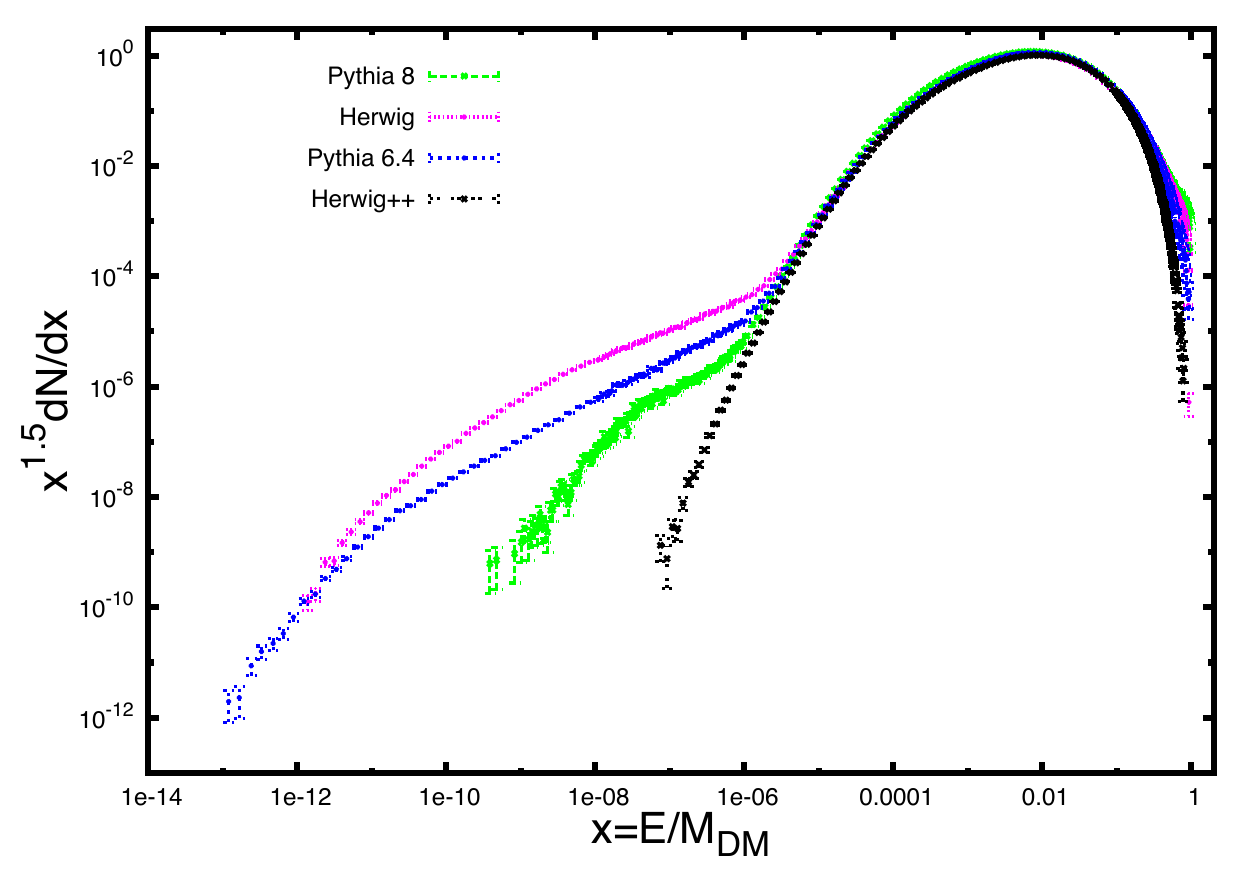}\includegraphics[height=100pt]{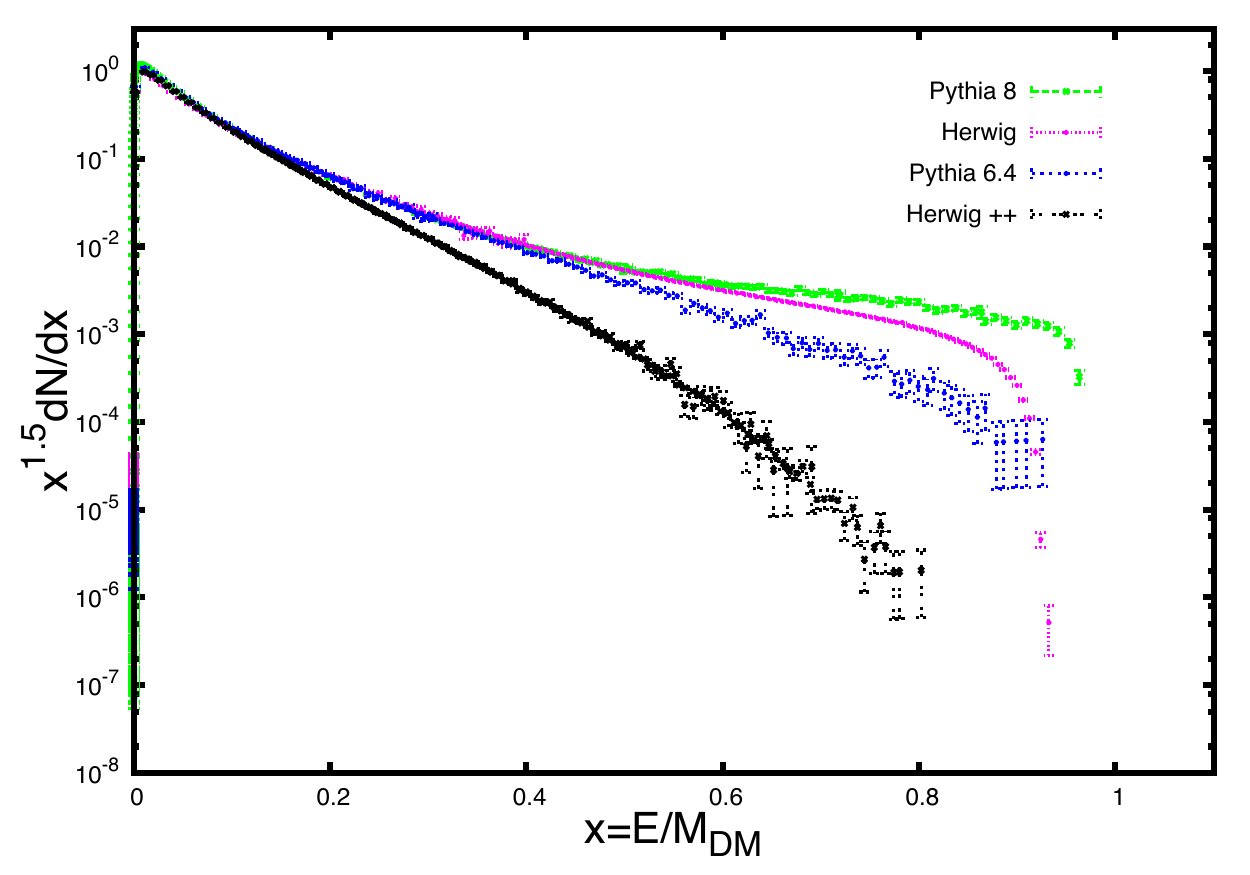}}
\caption{
{\it (Left--panel)} $t \bar t$ annihilation channel with $M_{\rm DM}=1$ TeV in logarithmic scale. At low energy the simulations are consistent down to $x\simeq10^{-6}$. \herwigpp drops down at $x\simeq10^{-7}$ and \pythia 8 does at $x\simeq 10^{-10}$, producing a higher number of photons that is 100 times higher than \herwigpp at $x\simeq 10^{-7}$, and almost 10 times lower than \pythia 6.4 at the same value of $x$. \pythia 6.4 cuts-off at $x\simeq 10^{-13}$ whereas \herwig does at $x\simeq 10^{-12}$ where
the two spectra match. For higher energies, \herwig provides a higher flux with a maximum factor of ten at $x\simeq 10^{-8}$.
{\it (Right--panel)} $t \bar t$ annihilation channel with $M_{\rm DM}=1$ TeV in linear scale. The four simulations are all manifestly inconsistent between them at very high energy. \herwigpp flux becomes lower from $x\simeq0.2$ onwards and cuts-off at $x<0.8$. At $x\simeq0.4$, \pythia 6.4 splits from \herwig and \pythia 8 that remain with higher flux.
\pythia 6.4 cuts-off before reaching $x=1$, such as \herwig does, although with very different spectral form and a separation of a factor ten at $x\simeq 0.8$. Finally, \herwig also splits from \pythia 8 at $x\simeq0.6$, producing the highest flux with cut-off at $x=1$. }
\label{fig:top1t}
\end{figure}


\subsection{Gamma--ray spectra from DM annihilation: $t\bar t$ channel}

The most remarkable differences between the four simulations packages appear in the $t\bar t$ channel.
To enable top decays in \pythia 6.4, the subroutine {\tt PYINIT()} has to be executed.
Alternatively, this process can be implemented by its dominant SM decay, i.e.
$t\rightarrow W^{+}b$ (or equivalently $\bar{t}\rightarrow W^{-}\bar{b}$)~\cite{Ce10}.
In order to maintain any non-perturbative effect, the initial state was made of
a four-particle state composed by $W^{+} b$ coming
from the $t$ quark and  $W^{-} \bar{b}$ from $\bar{t}$ anti-quark. These choices conserve all
kinematics and color properties from the original pair and show the same results as the {\tt PYINIT()} case.
Starting from this configuration, the authors forced decays and
hadronization processes to evolve as \pythia does.
Therefore, the gamma--rays spectra corresponding to this channel
have also been included for \pythia 6.4 in our analysis using this procedure.
For this channel we have studied two DM masses $500$ GeV and 1 TeV.
The simulated spectra appear very similar in the range $10^{-5}<x<0.1$.
Nonetheless, at lower and higher energies the four are quite different.
At large energies, \pythia 8 gives the highest flux being able to acquire non-null flux for $x\approx 1$. The smallest flux is again for \herwigpp whereas \pythia 6.4 and \herwig lie in between the other two. These facts can be seen in Figs. \ref{fig:top5g} and \ref{fig:top1t}.
The four spectra also differ at high energy due
to the (absence of) implementation of Bremsstrahlung effects. All the possibilities were summarized in Table \ref{topCodesTab}.
At low energy the differences may be associated as in the $\tau^{+}\tau^{-}$ both to the cut-off in the lowest energy allowed for photons
and to the presence or not of the \texttt{QEDRadiationHandler} in the simulation.\\

\begin{table}[tb]
\centering
\begin{tabular}{|c|c|}
\hline
\hline
Package &Bremsstrahlung\\
\hline
\hline
\pythia 6.4  & Implemented\\
\hline
\pythia 8  & Implemented\\
\hline
\herwig &  Partially implemented\\
\hline
\herwigpp  & Not implemented\\
\hline
\end{tabular}
\caption{
Simulations are strongly affected by the inclusion of Bremsstrahlung radiation and consequently the spectra turn out to look very different at high energy.}
\label{topCodesTab}
\end{table}


\section{Implications to WIMPs phenomenology}
\label{4}
Monte Carlo generators are essential tools for indirect searches of dark matter.
The simulated spectra generated by \pythia 6.4, \pythia 8, \herwig and \herwigpp allow to get predictions about the signal coming from DM annihilation and/or decay.
The choice of the Monte Carlo generator software may affect the predictions on both constraints and upper/lower limits to be imposed on DM annihilation cross section, relic density, astrophysical factor and other relevant quantities.
As we discussed in the previous sections, the gamma--ray spectra appear more similar at the energy corresponding to the peak of emission, but important differences appear at lower and higher energies.
Lower energies are less important in the context of indirect searches, because of the dominance of astrophysical background components.
However, the spectra at high energies could be of some interest.
As an illustrative example, the next Cherenkov Telescope Array (CTA) is expected to extend the accessible energy range from well below 100 GeV to above 100 TeV~\cite{CTA} and therefore may cover a wide range of high gamma--ray energies and signatures of DM annihilation in a wider range of masses than for instance FERMI-LAT satellite. \\
%

\begin{figure}[tb]
\centering
\subfigure[$W^+W^-$ channel]{
   \includegraphics[width=0.48\columnwidth] {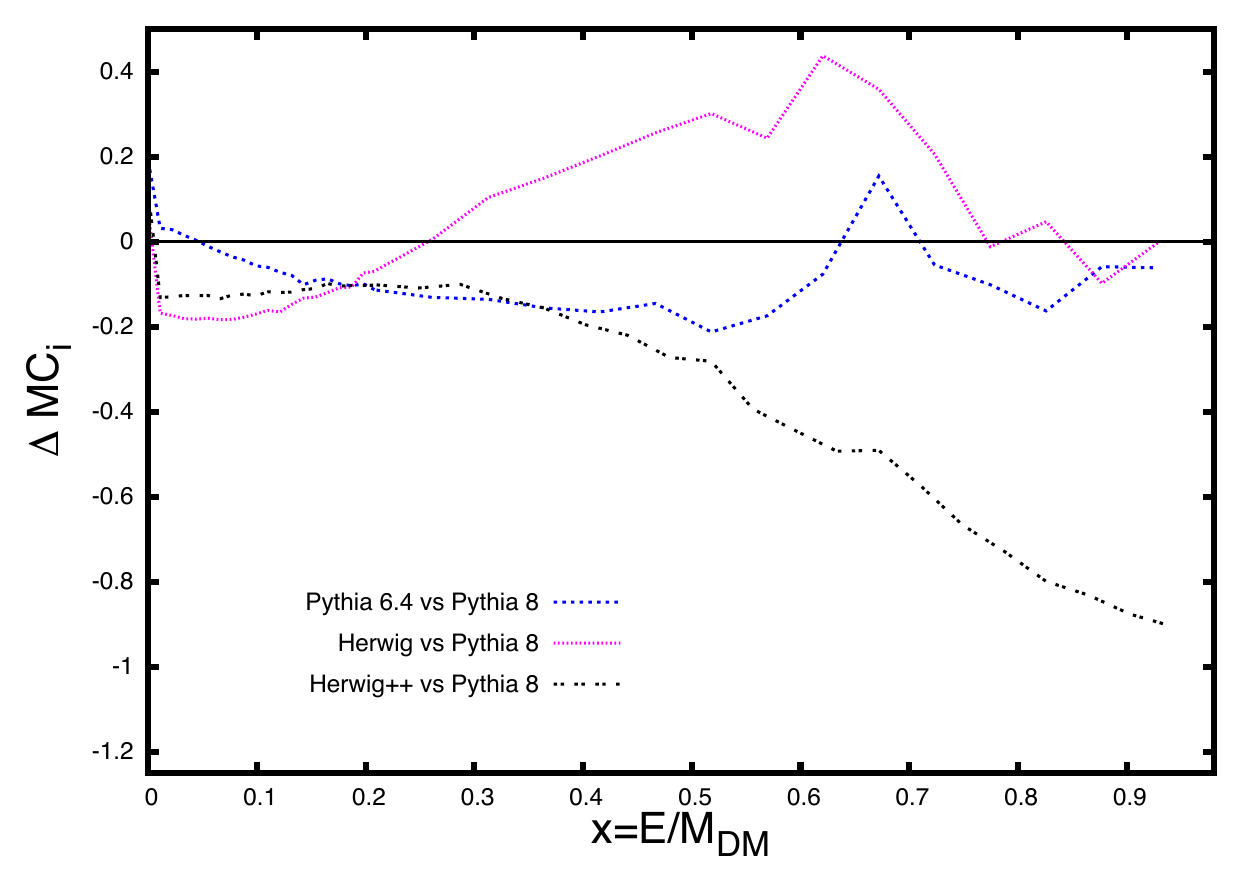}
   \label{ER_w}
 }
\subfigure[$b\bar b$ channel]{
   \includegraphics[width=0.48\columnwidth] {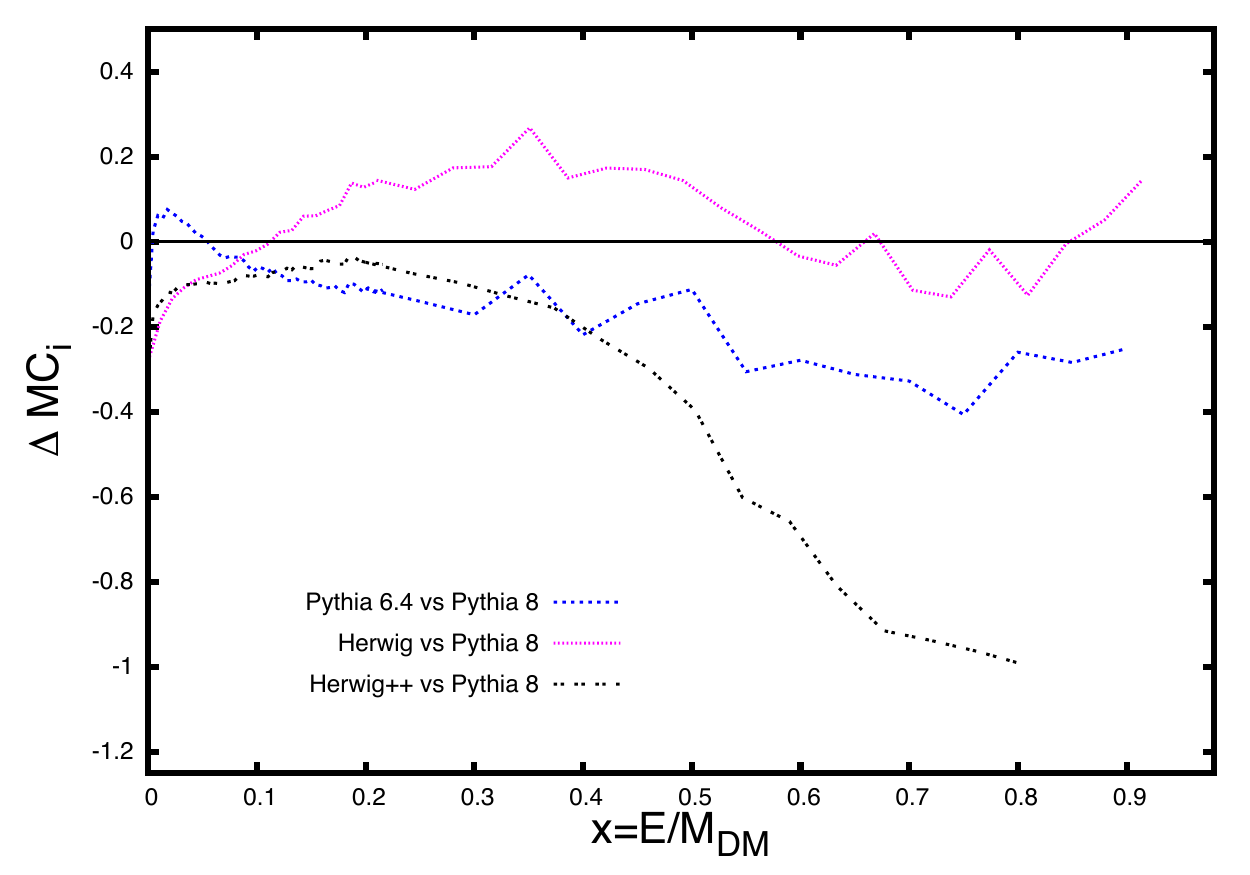}
   \label{ER_b}
 }
\subfigure[$\tau^+\tau^-$ channel]{
   \includegraphics[width=0.48\columnwidth] {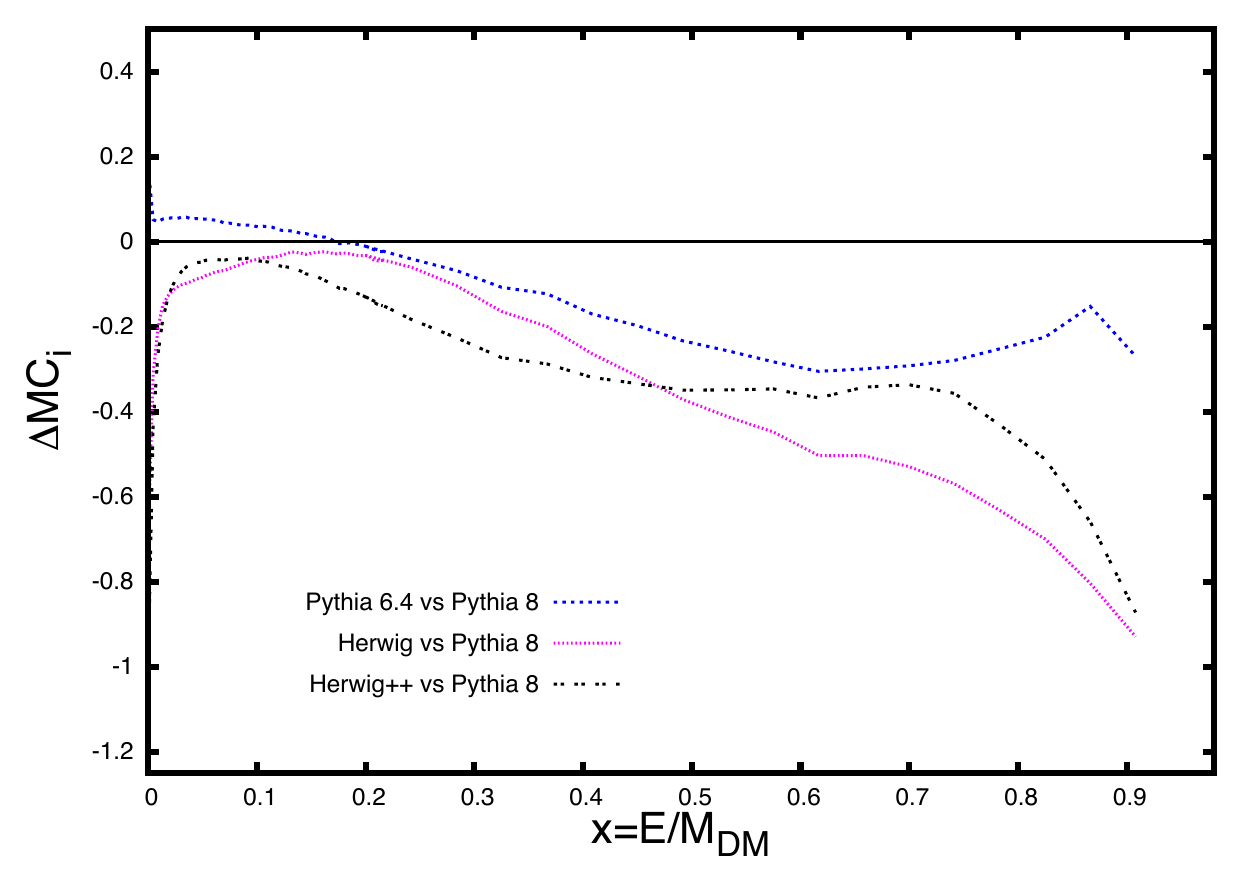}
   \label{ER_tau}
 }
\subfigure[$t\bar t$ channel]{
   \includegraphics[width=0.48\columnwidth] {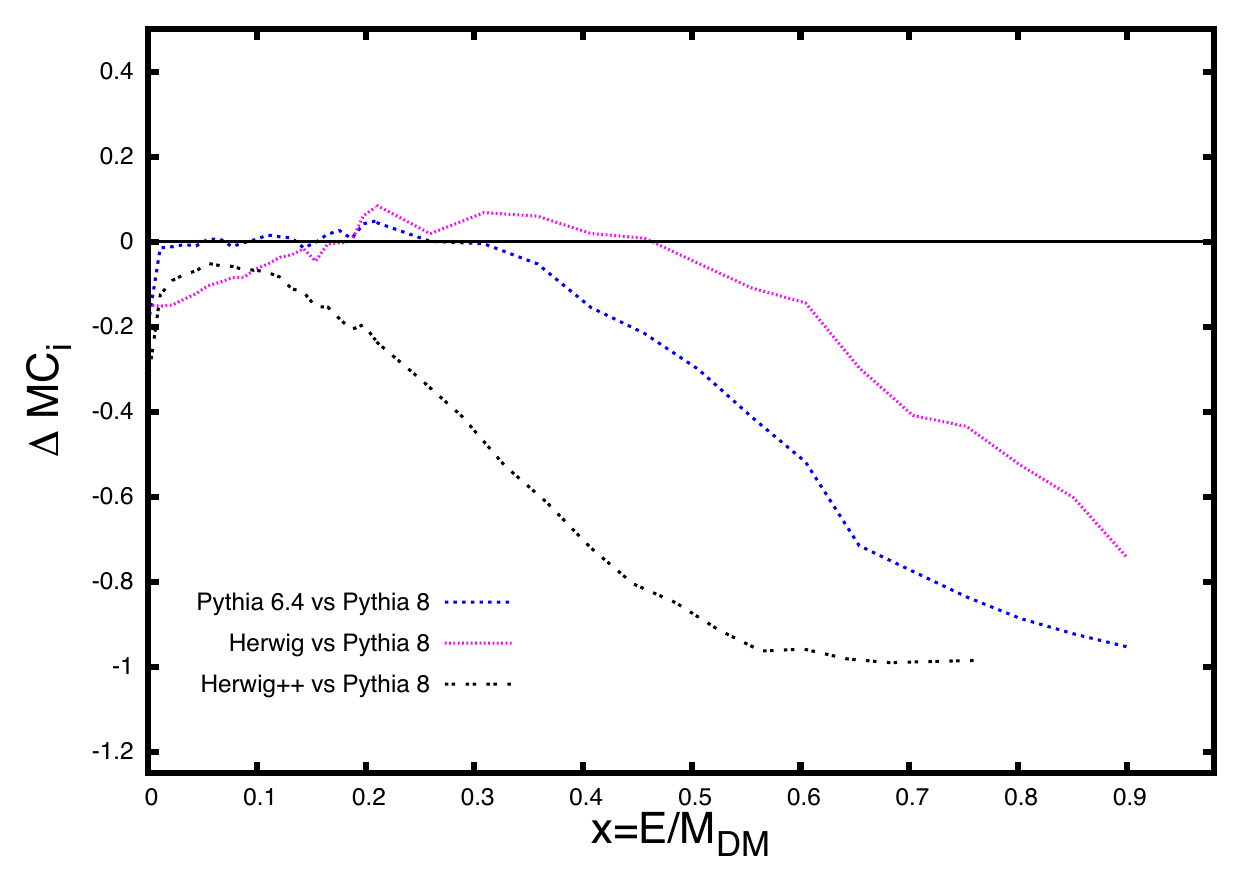}
   \label{ER_top}
 }
\caption{
Relative deviations versus $x$ at $M_{\rm DM}=1$ TeV.
The full horizontal line at zero represents \pythia 8.
The dashed blue line holds for \pythia 6.4 vs. \pythia 8,
the dotted one is \herwig Fortran vs. \pythia 8 and
the two-dotted one is \herwigpp vs. \pythia 8.
}
\label{ErrRel}
\end{figure}

Since \pythia 8 includes both a good description of the $t$ quark behavior and the QED radiation, we use it to compare with the other generators.
%
We present the Monte Carlo relative deviation ($\Delta {\rm MC}_i$) with respect to \pythia 8 in Fig. \ref{ErrRel}, defined as
\begin{equation}
\Delta MC_i =
\;\frac{MC_i\, - {\rm PYTHIA} \, 8}
{{\rm PYTHIA} \, 8},
\label{RelErr}
\end{equation}
where $MC_i$ stands for \pythia 6.4, \herwig and \herwigpp.
For a DM mass of $1$ TeV, the relative deviations are always less than $20\%$ up to $x=0.2$. For the whole high energy range, \pythia 6.4 produces typically less photons with a maximum relative error of $50\%$ with respect to \pythia 8, apart from the $t \bar t $ channel for which the strong approximation leads to differences up to $100\%$.
\herwig exhibits deviations lower than $50\%$ for the $W^+W^-$ channel up to $x\simeq0.6$,
similar deviations are found for $b\bar b$ up to $x\simeq0.5$ and for almost all the high energy
range (up to $x=0.8$) for $\tau^+\tau^-$. In the case of  $t\bar t$ channel, deviations below $50\%$
are found  just below $x\simeq0.3$. \herwigpp shows differences up to $100\%$ for all the annihilation channels when the energy increases beyond those values.

On the other hand, the total number of photons produced by each event or multiplicity, also affects the constraints both in the sense of annihilations cross section and astrophysical factor.
In indirect searches a typical significance of the signal between $2\sigma$ and $5\sigma$ with respect to the background is demanded.
Apart from the specific characteristics of the detector, the flux of photons depends upon the DM density and the distance and distribution of the sources.
All these dependences are taken into account by the astrophysical factor $\langle J \rangle $ and the boost factor $b$.
Thus, two simulations should give different number of photons for the same number of events, this situation will affect the parameters $\langle J\rangle$ and $b$.

%

\begin{figure}[tb]
\centering
\subfigure[$W^+W^-$ channel.]{
   \includegraphics[width=0.48\columnwidth] {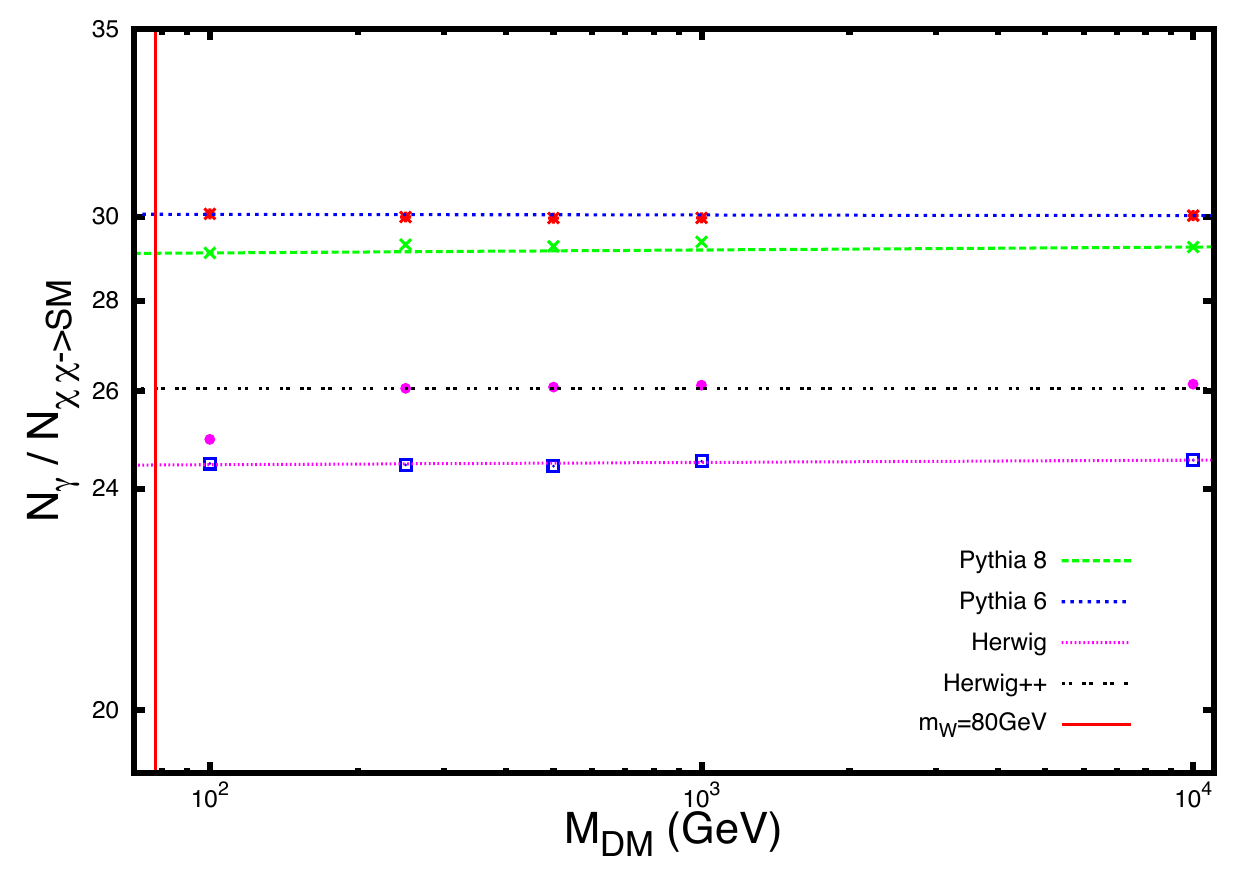}
   \label{mult_w}
 }
\subfigure[$b\bar b$ channel]{
   \includegraphics[width=0.48\columnwidth] {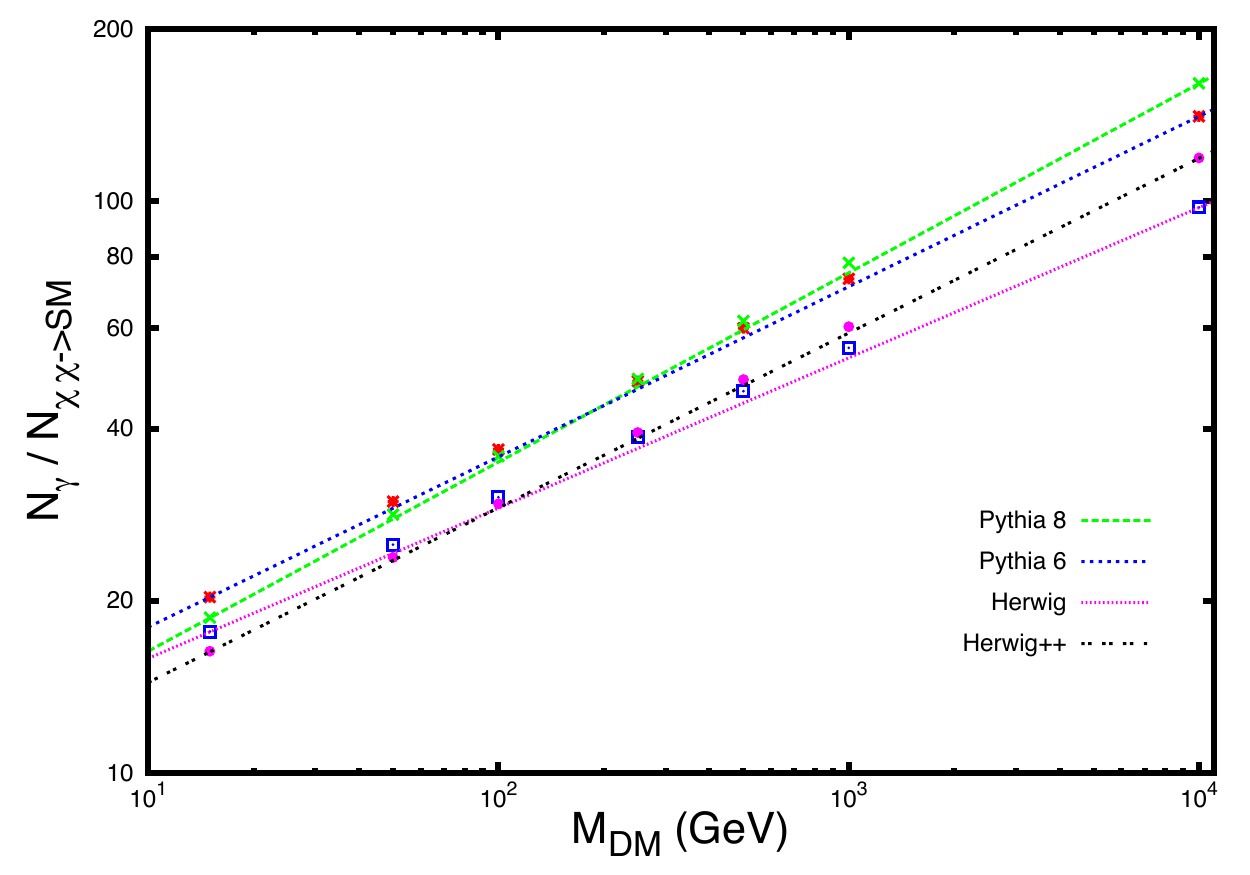}
   \label{mult_b}
 }

\subfigure[$\tau^+\tau^-$ channel]{
   \includegraphics[width=0.48\columnwidth] {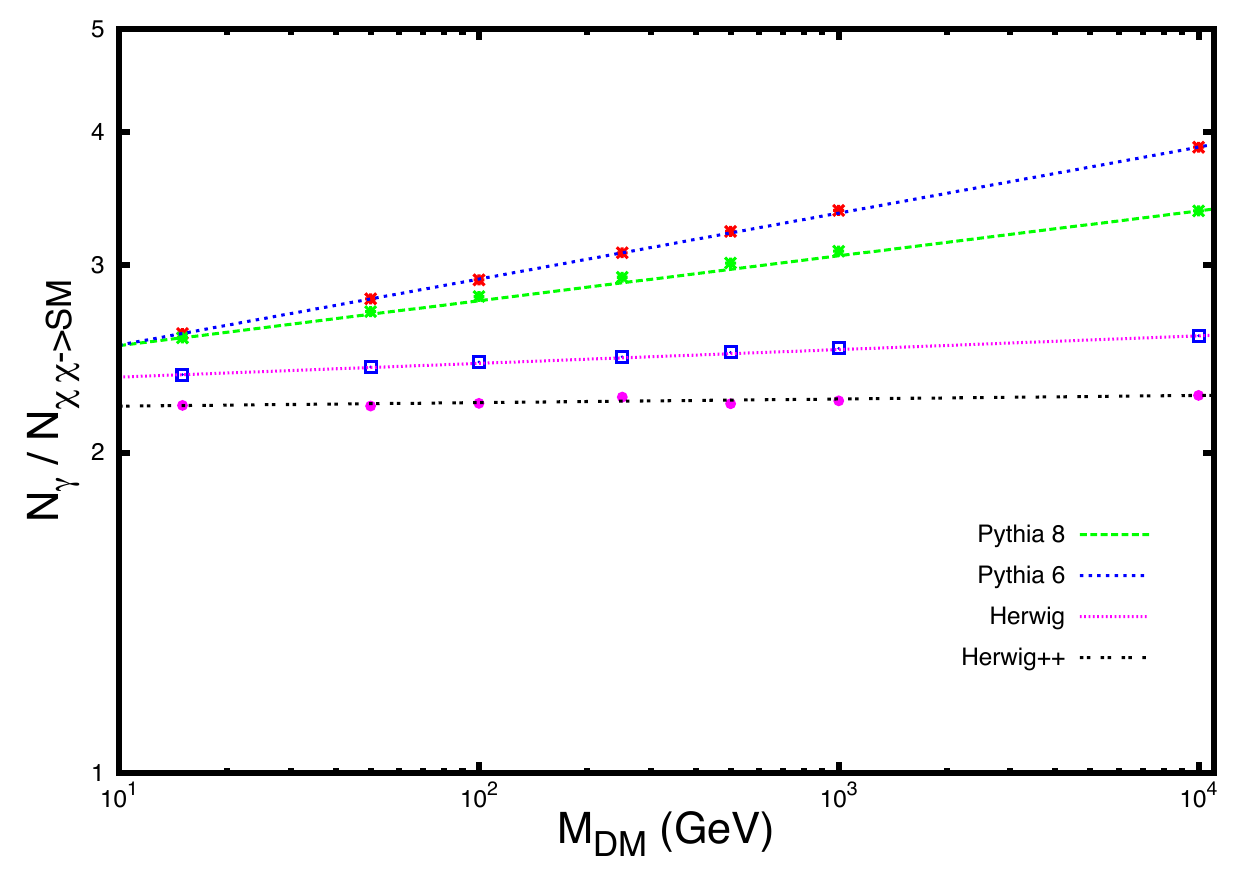}
   \label{mult_tau}
 }
\subfigure[$t\bar t$ channel]{
   \includegraphics[width=0.48\columnwidth] {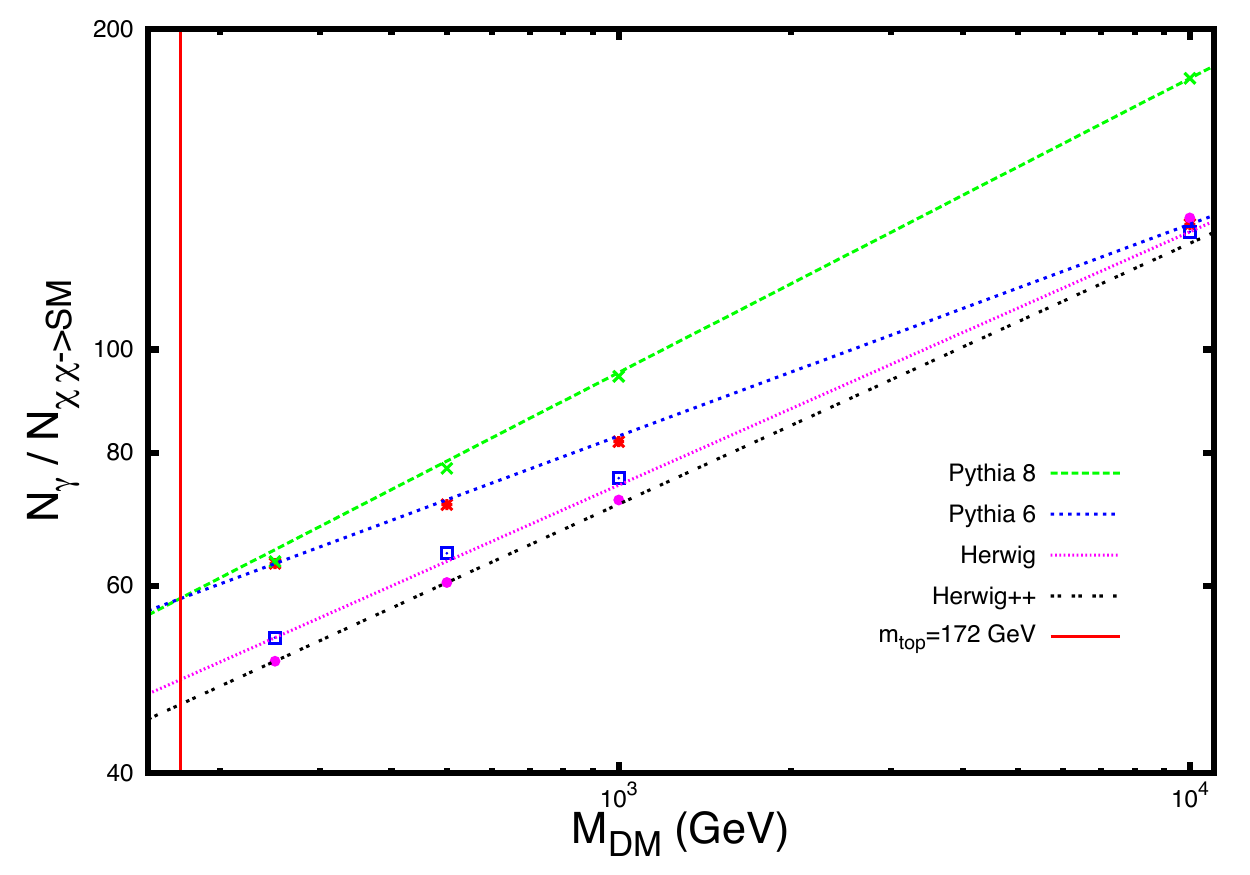}
   \label{mult_top}
 }
\caption{
Multiplicity of the four Monte Carlo generators for each annihilation channel.  $W^+W^-$ annihilation channel (upper left panel): Regardless the DM mass value, \pythia 6.4 provides the upper limit to the number of generated photons, while \herwig Fortran provides the lower limit with $23\%$ difference between them; $b\bar b$ annihilation channel (upper right panel): At $M_{\rm DM}\sim200$ GeV, the multiplicity of the two versions of \pythia is the same, as for the multiplicity of the \herwig versions, but different between them. For that value of the mass,
the relative deviation on multiplicity between \pythia and \herwig codes almost attains $100\%$; $\tau^+\tau^-$ annihilation channel (lower left panel): The maximum difference between the four simulations multiplicities ranges between $20\%$ at low energy up to $72\%$ at higher energy; $t\bar t$ annihilation channel (lower right panel): Relative deviations run from $20\%$ up to $30\%$ depending on the energy of the event.}
\label{Multi}
\end{figure}

As we can see in Fig. \ref{Multi}, the multiplicity depends not only on the Monte Carlo event generator, but also on the energy of the event and the annihilation channel. In this study, we set a lower photon energy cut-off of $x_C=10^{-5}$. It means that the energy cut-off increases with the DM mass. This kind of DM mass depending cut-off allows to reject photons of lower energies, where the simulations present important differences. However, the excluded range of the spectrum is not relevant for gamma-ray observations. This cut-off is also compatible with typical gamma-ray detectors energy thresholds. As an example, for a DM mass of $10$ TeV, the corresponding energy cut lies at $100$ MeV. Detector energy thresholds are typically around $1-10$ GeV depending on the particular experimental device \cite{branonsgamma}. In any case, we have checked that our results and conclusions about the different multiplicities do not depend on the particular choice of this cut-off. Thus we have tested the robustness of our analysis with $x_C=10^{-3}$ and $M_C=1$ GeV. In most of the cases \pythia 6.4 gives the multiplicity upper limit, except for the $t\bar t$ annihilation channel -- maybe due to the approximation of such process \cite{Ce10} -- and $b\bar b$ channel at the range  $M_{\rm DM}>200$ GeV. On the other hand, the lower limit is given by \herwigpp in most of the cases, except for $W^+W^-$ and $b\bar b$ (the last one, up to $M_{\rm DM}>200$ GeV) annihilation channel.

The multiplicity behavior is well approximated by the following power law relation with the DM mass:
\begin{equation}
\frac{N_{\gamma}}{N_{\chi\chi \rightarrow SM}}\simeq a\cdot \left(\frac{M}{1\,\text{GeV}}\right)^b\;,
\label{pl}
\end{equation}
where the $a$ and $b$ coefficients depend on both the Monte Carlo simulator and the annihilation channel. When the SM particle is fixed, cosmological constraints obtained by means of  the total number of generated gamma photons might depend on the Monte Carlo simulation. As in the previous analysis, in Table \ref{coeffMult} we give the relations between the total number of photons generated by \pythia 6.4, \herwig and \herwigpp with respect to \pythia 8.

\begin{table}[tb]
\centering
\begin{tabular}{|c|c|c|c|c|}
\hline
\hline
Software/\pythia 8 & $W^+W^-$ & $b\bar b$ & $\tau^+\tau^-$ & $t\bar t$\\
\hline
\hline
\multirow{2}{*}{\pythia 6.4} &$A=1.04$  & $A=1.18$&$A=0.96$&$A=1.49$\\
&$B=0$&$B=-0.033$&$\,\,B=0.020$&$B=-0.077$\\
\hline
\multirow{2}{*}{\herwig} & $A=0.84$ & $A=1.13$&$A=1.00$&$A=1.02$\\
&$\,\,B=0$&$B=-0.068$&$B=-0.029$&$B=-0.038$\\
\hline
\multirow{2}{*}{\herwigpp} & $A=0.90$ & $A=0.93$&$A=0.96$&$A=0.93$\\
&$B=0$&$B=-0.025$&$B=-0.039$&$B=-0.031$\\
\hline
\hline
\end{tabular}

\begin{tabular}{cc}
 & \\
\end{tabular}

\begin{tabular}{|c|c|c|c|c|}
\hline
\hline
\multirow{2}{*}{\pythia 8} &$a=28.9$  & $a=7.62$&$a=2.29$&$a=14.1$\\
&$b=0.001$&$b=0.331$&$\,\,b=0.042$&$b=0.276$\\
\hline
\hline
\end{tabular}
\caption{Relative behaviors in the total number of photons produced by \pythia 6.4, \herwig and \herwigpp with respect to \pythia 8 in the range $15$ GeV - $10$ TeV. Here $A=a_{MC_i}/a_{\text{\pythia 8}}$ and $B=b_{MC_i}-b_{\text{\pythia 8}}$. \pythia 8 multiplicity parameters are listed at the end of the Table.}
\label{coeffMult}
\end{table}

Let us summarize the situation as follows:\\

\begin{itemize}

\item \textbf{$W^+W^-$ annihilation channel}: Roughly speaking \pythia 6.4 generates one more photon than \pythia 8 for each event, while \herwigpp and \herwig Fortran produce 3 and 5 photons less, respectively. Above $\simeq200$ GeV, this fact introduces a deviation on the multiplicity of $\sim4\%$ between \pythia 6.4 and \pythia 8, of $\sim16\%$ between \herwig and \pythia 8 and of $\sim10\%$ between \herwigpp and \pythia 8. Between \pythia 6.4 and \herwig in Fortan and \herwigpp the deviation is $\sim23\%$ and $\sim15\%$ , respectively.  Finally, the deviation between \herwig and \herwigpp is $\sim6\%$. For kinematic reasons, no photons are produced at energies lower than the mass of the W boson, that is the reason of the cut around $\simeq80$ GeV.

\item \textbf{$b\bar b$ annihilation channel}: At $M_{DM} \simeq200$ GeV, the deviation between the multiplicity of the two versions of \pythia is the less than $1\%$, as for at $100$ GeV and \herwig versions, but different between them. At $150$ GeV, the number of photons produced by the Fortran versions of \pythia code is a $22\%$ bigger than the \herwig one. For masses below $\sim200$ GeV the upper limit is given by \pythia 6.4 whereas the lower one is provided by \herwigpp. At $10$ TeV the deviation reach the maximum value of $\sim13\%$, $\sim40\%$ and $\sim26\%$ between \pythia 6.4, \herwig, \herwigpp and \pythia 8, respectively.
On the other hand, at $M>200$ GeV \pythia 8 gives the upper limit and \herwig Fortran the lower one.

\item \textbf{$\tau^+\tau^-$ annihilation channel}: The number of photons per event produced by the four Monte Carlo generators is very similar for this channel, but very low. This fact introduce a very important difference in percent, that reach the maximum of $\sim42\%$ at $10$ TeV between \pythia 6.4 and \herwigpp. \pythia 6.4 gives here the upper limit, followed by \pythia 8, \herwig Fortran and \herwigpp. At lower energies the difference between upper and lower limit is less than $20\%$, and increase up to $72\%$ at higher DM mass.

\item \textbf{$t\bar t$ annihilation channel}: As in the case of $W^+W^-$ channel, no photons are produced at energies lower than the mass of the top quark because of kinematic reason. Always \pythia 8 gives here the upper limit, followed by \pythia 6.4, \herwig Fortran and \herwigpp. All the multiplicities depend on the DM mass in a exponential way, but with different exponents. At lower energies the deviation between upper and lower limits is about $20\%$, and around $30\%$ for events at higher energies.


\end{itemize}


\section{Conclusions}
\label{5}

We have analyzed the gamma--ray spectra produced by four Monte Carlo event generator software, namely \pythia 6.4, \pythia 8, \herwig Fortran and \herwigpp. These spectra have been largely used in the framework of dark matter indirect searches and the differences between them may affect the results for those investigations. Although gamma--ray spectra have been generated for dark matter annihilating in all possible quark-antiquark, leptonic and bosons channels, we chose to show a representative sample of them ($b\bar b$ for the quark-antiquark case, $\tau^+\tau^-$ for the leptonic one and $W^+W^-$ boson annihilation channels). We also included the particular case of the $t\bar t$ and studied it separately.

At the energy of maximum flux, where the simulations are well fitted to LEP or LHC data, the differences between packages are less than $20\%$. This statement is always true in the range $0.01<x<0.2$ with possible extension of the range depending on the annihilation channel and the energy of the event (see the bulk of this communication for further details).  On the one hand, at lower energy the spectra appear very different between them, depending strongly on the cut-off set for the minimal allowed energy in the parton shower.  On the other hand, differences also appear at higher energy. For all the studied channels, the implementation absence of Bremsstrahlung radiation generated by high energy leptons in \herwigpp leads to a smaller number of high-energy photons when compared to the other softwares. Moreover, in the case of the $t\bar t$ annihilation channel, there is an additional effect due to the fact that the top quark behavior phenomenology has been improved in the codes released in the last years.
%
Thus, whereas for \pythia 6.4 this channel was approximated through the decay
into $W$ and $b$, higher order effects have been included in the newest software generations. Due to the combination of these two factors, we conclude that
the most reliable Monte Carlo event generator software for gamma--ray spectra is \pythia 8. For this reason we got estimations for the relative deviations for \pythia 6.4, \herwig Fortran and \herwigpp with respect to \pythia 8.

We conclude that further implementation is needed in \herwigpp in order to improve its competitiveness in the gamma--ray sector. For the other three Monte Carlo event generators under study in this work, the gamma--ray spectra simulated show also important differences. Without taking into account very low energies, the relative deviations can only be bounded by $50\%$ for the hadronic ($b\bar b$) and electro-weak channels ($W^+W^-$). The situation for
the $t\bar t$ channel and the leptonic ones ($\tau^+\tau^-$) is even worse. At high energies, the discrepancies can reach $100\%$. In fact, the photon fluxes predicted by the different generators can differ in several orders of magnitude.
On the other hand, the situation for the total number of produced photons improves a little, and the maximum difference is a factor $2$ within the studied mass region.

These significative differences can play an important role in misunderstanding dark matter signatures. For example, in a dark matter study, once the astrophysical factor is obtained by fitting the dark matter gamma--ray spectra, these discrepancies may introduce a deviation on the boost factor proportional to the difference on the multiplicity. This effect can be easily estimated with the help of Eq. (\ref{pl}) and Table \ref{coeffMult}. However, we have shown that the simulated spectral shapes can be very different and this fact may have a large impact in the analysis.

\acknowledgments
This work was supported by UCM FPI grants G/640/400/8000 (2011 Program), the Spanish MINECO 
projects numbers FIS2011-23000, FPA2011-27853-C02-01, FPA2011-22975 and  MULTIDARK CSD2009-00064 (Consolider-Ingenio 2010 Programme).
AdlCD also acknowledges financial support from Marie Curie - Beatriu de Pin\'os contract BP-B00195, Generalitat de Catalunya and ACGC, University of Cape Town.
RL also acknowledges  Prometeo/2009/091 (Generalitat Valenciana) and 
EU ITN UNILHC PITN-GA-2009-237920 financial support.

\bibliographystyle{JHEP}

\begin{thebibliography}{99}

\bibitem{DM} L.~Covi, J.~E.~Kim and L.~Roszkowski, Phys.\ Rev.\ Lett.\  {\bf 82}, 4180 (1999);
  J.~L.~Feng, A.~Rajaraman and F.~Takayama, Phys.\ Rev.\ D {\bf 68}, 085018 (2003); 
  J.~L.~Feng, A.~Rajaraman and F.~Takayama,  Int.\ J.\ Mod.\ Phys.\ D {\bf 13}, 2355 (2004); 
  J.~A.~R.~Cembranos, J.~L.~Feng, A.~Rajaraman and F.~Takayama, Phys.\ Rev.\ Lett.\  {\bf 95}, 181301 (2005); 
  J.~A.~R. Cembranos, J.~L.~Feng, L.~E.~Strigari,  Phys.\ Rev.\  D {\bf 75}, 036004 (2007); 
  J.~A.~R.~Cembranos, Phys.\ Rev.\ Lett.\  {\bf 102}, 141301 (2009); 
  Phys.\ Rev.\  D {\bf 73}, 064029 (2006); 
  J.~A.~R.~Cembranos, J.~L.~Diaz-Cruz and L.~Prado, Phys.\ Rev.\ D {\bf 84}, 083522 (2011). 

\bibitem{WIMPs}
  H.~Goldberg, Phys.\ Rev.\ Lett.\  {\bf 50}, 1419 (1983); 
  J.~R.~Ellis {\it et al.}, Nucl.\ Phys.\ B {\bf 238}, 453 (1984); 
  K. Griest and M. Kamionkowski, Phys. Rep. \textbf{333}, 167 (2000);
  J.~A.~R.~Cembranos, A.~Dobado and A.~L.~Maroto,  Phys.\ Rev.\ Lett.\  {\bf 90}, 241301 (2003); 
  Phys.\ Rev.\ D {\bf 68}, 103505 (2003); 
  Phys.\ Rev.\ D {\bf 73}, 035008 (2006); 
  Phys.\ Rev.\ D {\bf 73}, 057303 (2006); 
  A.~L.~Maroto, Phys.\ Rev.\ D {\bf 69}, 043509 (2004); 
  Phys.\ Rev.\ D {\bf 69}, 101304 (2004); 
  A. Dobado and A. L. Maroto, Nucl. Phys. B \textbf{592}, 203 (2001); 
  Int. J. Mod. Phys. {\bf D13}, 2275 (2004) [hep-ph/0405165]; 
  J.~A.~R.~Cembranos  {\it et al.}, JCAP {\bf 0810}, 039 (2008). 

\bibitem{Coll}
  J.~Alcaraz {\it et al.}, Phys. Rev.{\bf D67}, 075010 (2003); 
  P. Achard {\it et al.}, Phys. Lett. {\bf B597}, 145 (2004); 
  Europhys.\ Lett.\  {\bf 82}, 21001 (2008); 
  J.~A.~R.~Cembranos, A.~Dobado and A.~L.~Maroto, Phys. Rev. {\bf D65} 026005 (2002); 
  J.\ Phys.\ A  {\bf 40}, 6631 (2007); 
  Phys. Rev. {\bf D70}, 096001 (2004); 
  J.~A.~R.~Cembranos {\it et al.}, AIP Conf.\ Proc.\  {\bf 903}, 591 (2007). 

\bibitem{isearches}
  J.~A.~R.~Cembranos and L.~E.~Strigari, Phys.\ Rev.\  D {\bf 77}, 123519 (2008); 
  J.~A.~R.~Cembranos, J.~L.~Feng and L.~E.~Strigari, Phys.\ Rev.\ Lett.\  {\bf 99}, 191301 (2007). 

\bibitem{simu}
  M. Cirelli et al.
  JCAP {\bf 1103}, 051 (2011) [Erratum-ibid.\  {\bf 1210}, E01 (2012)].

\bibitem{Cembranos:2012nj}
  J.~A.~R.~Cembranos, V.~Gammaldi and A.~L.~Maroto,
  Phys.\ Rev.\ D {\bf 86}, 103506 (2012); 
  arXiv:1302.6871 [astro-ph.CO]. 

\bibitem{branonsgamma}
  J.~A.~R.~Cembranos, A.~de la Cruz-Dombriz, V.~Gammaldi and A.~L.~Maroto,
  Phys.\ Rev.\ D {\bf 85}, 043505 (2012). 

\bibitem{Ce10}
  J.~A.~R.~Cembranos, A.~de la Cruz-Dombriz, A.~Dobado, R.~Lineros and A.~L.~Maroto,
  Phys.\ Rev.\  D {\bf 83}, 083507 (2011); 
  AIP Conf.\ Proc.\  {\bf 1343}, 595-597 (2011);
  J.\ Phys.\ Conf.\ Ser.\  {\bf 314}, 012063 (2011);
  A.~de la Cruz-Dombriz and V.~Gammaldi, arXiv:1109.5027 [hep-ph];
  http://teorica.fis.ucm.es/PaginaWeb/photon\_spectra.html

\bibitem{HessGC}
  A.V. Belikov, G. Zaharijas, J. Silk,  Phys. Rev. D 86, 083516 (2012).

\bibitem{Fermi}
  A. A. Abdo et al. [arXiv:astro-ph.CO/1001.4531v1] (2010).
  M. Chernyakova {\em et.~al.}, ApJ {\bf 726}, 60 (2011);
  T.~Linden, E.~Lovegrove and S.~Profumo,  arXiv:1203.3539 [astro-ph.HE].

\bibitem{CANG}
  K. Tsuchiya, R. Enomoto, L. T. Ksenofontov et al. ApJ, 606, L115 (2004).

\bibitem{VER}
  K. Kosak, H. M. Badran, I. H. Bond et al., ApJ, 608, L97 (2004).

\bibitem{Aha}
  F. Aharonian, A. G. Akhperjanian, K.M. Aye et al. A\&A, 425, L13 (2004b).

\bibitem{HESS}
  F. Aharonian, A. G. Akhperjanian, K.M. Aye et al. A\&A, 503, 817 (2009).

\bibitem{MAG}
  J. Albert, E. Aliu, H. Anderhub et al., ApJ, 638, L101 (2006).
  MAGIC collaboration, [arXiv:astro-ph/1103.0477v1] (2011).

\bibitem{WMAP}
  E.~Komatsu {\it et al.} [WMAP Collaboration],  ApJ. Suppl. 192 18 (2011).

\bibitem{CTA}
  Maier G. [arXiv:astro-ph.IM/0907.5118v1] ; The CTA consortium [arXiv:astro-ph.IM/1003.3703v2]; http://www.cta-observatory.org/

\bibitem{str}
  G.R. Blumenthal, S.M. Faber, R. Flores, J. R. Primack,
  ApJ {\bf 301}, 27 (1986);
  O.~Y.~Gnedin, A.~V.~Kravtsov, A.~A.~Klypin and D.~Nagai,
  ApJ  {\bf 616}, 16 (2004);
  F.~Prada, A.~Klypin, J.~Flix Molina, M.~Mart\'inez, E.~Simonneau,
  Phys.\ Rev.\ Lett.\  {\bf 93}, 241301 (2004);
  E.~{Romano-D{\'{\i}}az}, I.~{Shlosman}, Y.~{Hoffman}, and C.~{Heller},
  ApJ {\bf 685}, L105 (2008); ApJ {\bf 702}, 1250  (2009);
  A.~V. Maccio' {\em et.~al.}, arXiv:1111.5620 [astro-ph.CO];   
  P.~Salucci, M.~I.~Wilkinson, M.~G.~Walker, G.~F.~Gilmore, E.~K.~Grebel, A.~Koch, C.~F.~Martins and R.~F.~G.~Wyse,
  arXiv:1111.1165 [astro-ph.CO];
  M.~Baldi and P.~Salucci,
  JCAP {\bf 1202}, 014 (2012); 
  G.~Castignani, N.~Frusciante, D.~Vernieri and P.~Salucci,
  Natural Sci.\  {\bf 4}, 265 (2012)
  I.~Cholis and P.~Salucci,
  Phys.\ Rev.\ D {\bf 86}, 023528 (2012). 

\bibitem{Seymour:2013ega}
  M.~H.~Seymour and M.~Marx,  
  arXiv:1304.6677 [hep-ph]. 

\bibitem{Beringer:1900zz}
  J.~Beringer {\it et al.}  [Particle Data Group Collaboration],
  Phys.\ Rev.\ D {\bf 86}, 010001 (2012). 

\bibitem{Altarelli:1977zs}
  G.~Altarelli and G.~Parisi,
  Nucl.\ Phys.\ B {\bf 126}, 298 (1977).  

\bibitem{81} G. Marchesini and B.R. Webber, Nucl. Phys. B {\bf 238} 1 (1984); Nucl. Phys. B {\bf 310} 461 (1988).

\bibitem{83}  T. Sj\"ostrand and P. Skands, Eur. Phys. J. C {\bf 39} 129 (2005).

\bibitem{CMW}
Catani, Webber, Marchesini, Nucl. Phys. B349 (1991) 635-654.

\bibitem{Py6}
  T. Sj\"ostrand, S. Mrenna, P. Skands, arXiv:0603175 [hep-ph].

\bibitem{Py8}
  T. Sj\"ostrand, S. Mrenna, P. Skands, arXiv:0710.3820v1 [hep-ph]; http://home.thep.lu.se/~torbjorn/pythia81.html

\bibitem{Her}
  G. Corcella {\it et al.}, arXiv:0011363v3 [hep-ph].

\bibitem{Her++}
  M. B\"ahr {\it et al.}, arXiv:0803.0883v3 [hep-ph]. S. Gieseke et al., arXiv:1102.1672v1 [hep-ph], K. Arnold et al., arXiv:1205.4902v1 [hep-ph].

\end{thebibliography}

\end{document}